\documentclass[prd,aps,a4paper,superscriptaddress,twocolumn,nofootinbib]{revtex4}
\usepackage{graphicx}
\usepackage{color}
\usepackage{dcolumn}
\usepackage{bm}
\usepackage{slashed}
\usepackage{amsmath}
\usepackage{latexsym}
\usepackage{amssymb}
\usepackage{mathrsfs}
\usepackage{amsfonts}
\usepackage{url}
\allowdisplaybreaks
\begin{document}
\title{Lorentz transformation of three dimensional gravitational wave tensor}

\author{Xiaokai He}
\affiliation{School of Mathematics and
Statistics, Hunan First Normal University, Changsha
410205, China}
\author{Xiaolin Liu}
\affiliation{Department of Astronomy, Beijing Normal University,
Beijing 100875, China}
\author{Zhoujian Cao
\footnote{corresponding author}} \email[Zhoujian Cao: ]{zjcao@amt.ac.cn}
\affiliation{Institute for Frontiers in Astronomy and Astrophysics, Beijing Normal University, Beijing 102206, China}
\affiliation{Department of Astronomy, Beijing Normal University, Beijing 100875, China}
\affiliation{School of Fundamental Physics and Mathematical Sciences, Hangzhou Institute for Advanced Study, UCAS, Hangzhou 310024, China}

\begin{abstract}
Recently there are more and more interest on the gravitational wave of moving sources. This introduces a Lorentz transformation problem of gravitational wave. Although Bondi-Metzner-Sachs (BMS) theory has in principle already included the Lorentz transformation of gravitational wave, the transformation of the three dimensional gravitational wave tensor has not been explicitly calculated before. Within four dimensional spacetime, gravitational wave have property of `boost weight zero' and `spin weight 2'. This fact makes the Lorentz transformation of gravitational wave difficult to understand. In the current paper we adopt the traditional three dimensional tensor description of gravitational wave. Such a transverse-traceless tensor describes the gravitational wave freedom directly. We derive the explicit Lorentz transformation of the gravitational wave tensor. The transformation is similar to the Lorentz transformation for electric field vector and magnetic field vector which are three dimensional vectors. Based on the deduced Lorentz transformation of the gravitational wave three dimensional tensor, we can construct the gravitational waveform of moving source with any speed if only the waveform of the corresponding rest waveform is given. As an example, we apply
our method to the effect of kick velocity of binary black hole. The adjusted
waveform by the kick velocity is presented.
\end{abstract}

\maketitle

\section{Introduction}
Since the first detection of gravitational wave (GW) by LIGO in 2015, the gravitational wave astronomy developed very quickly. Binary black holes, binary neutron stars and neutron star-black hole binaries have been found. People traditionally deem that binary compact objects form in two possible channels including isolated evolution which happens in field \cite{2016Natur.534..512B} and dynamical encounter which happens in clusters \cite{Antonini2016MERGING,2016ApJ...824L...8R}. Many binary black holes found by GW detection are much more massive than people ever expected \cite{PhysRevLett.125.101102}. Such finding stimulated discussions and studies of the formation problem of such binary systems. Recently, people propose one new channel for the binary formation. These GW binaries may form in the accretion disk of a suppermassive black hole \cite{Bartos2017Rapid,2017MNRAS.464..946S,PhysRevLett.123.181101,McKernan2019Ram,PhysRevLett.124.251102}. The authors of \cite{Bellovary2016Migration} found that the migration traps of the accretion disk may make the binary locate at the traps. If the disk is thick, the pressure gradient may change the structure of the migration traps \cite{2021MNRAS.505.1324P} which results in a trap locating at the distance of several gravitational radius of the central black hole. How do the binary black holes (BBH) detected by gravitational waveform has become a very interesting problem.

The binary black hole formed near a supermassive black hole will be affected by the gravitational potential of the central black hole \cite{2019MNRAS.485L.141C,PhysRevD.101.063002,PhysRevLett.126.021101,PhysRevLett.126.101105,PhysRevD.103.124044,Li2021Orbita,Fang2019Impact,PhysRevD.101.083031}. One of such effects is that the binary's barycentre will move respect to the detector. Here we care about the problem how such motion may change the waveform radiated by the binary. In addition to the Doppler shift \cite{PhysRevLett.117.011101,10.1093.mnras.stz2077}, other corrections of the waveform may be introduced by the relative motion between the source and the detector \cite{PhysRevD.100.063012,PhysRevD.101.083028,PhysRevLett.127.041102}. If ones can determine the moving velocity of the gravitational wave source \cite{PhysRevD.104.123025,2021arXiv211206425T}, such information will be helpful to distinguish the formation channel of the BBHs.

Besides the effect of a central supermassive black hole, the velocity dispersion of galaxy clusters may also provide a relative motion between binary black hole and the GW detector \cite{PhysRevLett.127.041102}. When the relative speed is slow, small velocity approximation can be used to treat the waveform changing problem \cite{PhysRevD.100.063012,PhysRevD.101.083028,PhysRevLett.127.041102}. Such small velocity condition is valid for galaxy velocity dispersion and binary black hole locating more than tens of gravitational radius away from the center supermassive black hole. If the binary black hole locates very near to the supermassive black hole \cite{2021MNRAS.505.1324P}, small velocity approximation may break down. And exact Lorentz transformation of gravitational waveform is expected. In the current paper we will present such transformation explicitly and express it in an electromagnetic-wave-like manner.

When considering the Lorentz transformation of gravitational wave, ones may correspondingly ask the tensor rank of gravitational wave. Unfortunately gravitational wave admits both `boost weight zero' and `spin weight 2' properties \cite{thorne88} which mean that gravitational wave behaves like both a scalar and a rank-two tensor. Essentially gravitational wave is neither a scalar nor a rank-two tensor. We need to rely on the Bondi-Metzner-Sachs (BMS) theory \cite{BonVanMet62,Sac62,PenRin88,he2015new,he2016asymptotical} to find out the Lorentz transformation of gravitational wave \cite{PhysRevD.93.084031}.

When the gravitational wave can be looked as a perturbation of the Minkowsky spacetime, the gravitational wave can be viewed as a rank-two tensor respect to the Lorentz group \cite{maggiore2008gravitational}. But the velocity involved in the Lorentz transformation can not be large, otherwise the perturbation condition of the rank-two tensor will break down. In addition, a transverse-traceless rank-two tensor will be transformed to a tensor which does not satisfy the transverse-traceless condition any more. Consequently people need to apply an additional transverse-traceless projection after the transformation.

People have already been used to describe gravitational wave with a three dimensional tensor, which is transverse-traceless. This three dimensional tensor is covariant respect to general three dimensional coordinate transformation. But it can not be treated in four dimensional viewpoint. This character is quite similar to that of an electric vector and a magnetic vector. Together with the Lorentz transformation we introduced in the current paper, the three dimensional tensor can describe gravitational wave as completely as the electric vector and the magnetic vector describing electromagnetic field. The Lorentz transformation does not change the transverse-traceless property of the gravitational wave tensor. Together with our Lorentz transformation rule, the three dimensional tensor provides a good tool to describe gravitational wave.

Actually, the BMS theory has already presented a BMS transformation of gravitational wave which includes Lorentz transformation, rotation transformation, translation transformation and even super-translation transformation in the four dimensional manifold view point \cite{PhysRevD.93.084031}. But such representation is quite hard for people who are not familiar with differential geometry. Especially usual astronomers will feel hard to understand such theory. This is very similar to the situation about electromagnetics in curved spacetime before the membrane paradigm proposed by Thorne and his coworkers in 1980s \cite{1986bhmp.book.....T}. At that time astronomers feel quite hard to understand the behavior of electromagnectis in curved spacetime although the four dimensional theory about such problem has been clear already. In contrast, the membrane paradigm uses the usual three dimensional language. Astronomers afterwards studied, applied and developed the electromagnetical theory in curved spacetime extensively. We hope the Lorentz transformation theory of three dimensional gravitational wave tensor presented in the current paper can play similar role as the membrane paradigm for gravitational wave astrophysics. Based on this Lorentz transformation theory of three dimensional gravitational wave tensor, astronomers can straightforwardly construct waveform model for kinds of moving sources if only the waveform of the corresponding rest waveform is known \cite{He_2022}.

The rest of this paper is arranged as following. We firstly review and comment the three dimensional tensor description of gravitational wave in the next section. Then we set up the Lorentz transformation relation based on BMS theory aiming to deduce the Lorentz transformation of gravitational wave in section III. Followed that we apply the BMS Lorentz transformation rule to electromagnetic wave in section IV. Along with the deducing of the Lorentz transformation rule for electromagnetic wave, we construct a key relation between two relative moving frames. Based on the BMS Lorentz transformation rule and the aforementioned key relation we constructed in section V the Lorentz transformation formula for gravitational wave tensor. In section VI, we
calculate the phase changing resulted from the Lorentz transformation based on the explicit
Lorentz transformation formula of a three dimensional transverse-traceless tensor. In section VII,  we explicitly construct the waveform for moving sources. The adjusted waveform by kick velocity of a BBH is presented there as an example of the waveform constructing process for moving sources.  At last  we give a summary and discussion in section VIII. 

Through the whole paper the units with $G=c=1$ are used. The Einstein sum rule is adopted. The indexes from $i$ to $n$ take values from 1 to 3. Other indexes take values from 0 to 3.

\section{Three dimensional tensor description of gravitational wave}
Essentially general relativity is a four dimensional theory. But three dimensional description can facilitate people to understand general relativity through traditional way. The membrane paradigm of black hole is a very good example of such object \cite{1986bhmp.book.....T}.

Physically gravitational wave admits two polarization modes which correspond to the two freedom of the gravitational wave. Consequently we can describe gravitational wave through a three dimensional tensor
\begin{align}
h_{ij}\equiv h_+ e_{ij}^++h_{\times}e_{ij}^\times,\label{eq22}
\end{align}
where $h_{+,\times}$ and $e_{ij}^{+,\times}$ are the two polarization modes and the corresponding bases. There is one and only one direction $\hat{N}^i$ (up to a sign) perpendicular to $h_{ij}$. Such direction indicates the propagating direction of the gravitational wave
\begin{align}
\hat{N}^ih_{ij}=0.
\end{align}

As a tensor, any coordinate including cartesian coordinate, spherical coordinate and others can be used to do the calculation. This is not new. Many literatures have already taken such facility \cite{maggiore2008gravitational,creighton2012gravitational}. Many astronomers have been familiar with the `transverse-traceless' property of GW which means just the above three dimensional tensor description. If in four dimensional viewpoint, many different descriptions may happen \cite{PenRin88,christodoulo2009,2021arXiv211206425T}.

But till now the above tensor description of gravitational wave is limited in three dimensional coordinate transformation. Physically it is limited to rotation transformations. Analogously, the electric vector and the magnetic vector are also just three dimensional tensor. But they can describe the four dimensional behavior of electromagnetic field quite well. The key point is there is a Lorentz transformation rule for the electric vector and the magnetic vector. To fill the gap of gravitational wave, we construct the Lorentz transformation rule for the gravitational wave tensor (\ref{eq22}) in the current paper. Equipped with the Lorentz transformation rule, the above three dimensional tensor description will be more powerful to study gravitational wave.
\section{Lorentz transformation within the BMS theory}
Within the BMS theory, the Lorentz transformation acted on two asymptotic inertial frames $(t,x,y,z)$ and $(t',x',y',z')$ can be expressed as \cite{PenRin88}
\begin{align}
\begin{pmatrix} t'+z'&x'+iy'\\x'-iy'&t'-z'\end{pmatrix}&=L\begin{pmatrix}t+z&x+iy\\x-iy&t-z\end{pmatrix}L^\dag,\label{eqII2}
\end{align}
where $\dag$ means transpose and complex conjugate (hermitian conjugate), and $L$ is a $2\times2$ hermitian matrix representing the Lorentz transformation. Corresponding to boost with relative velocity $\vec{v}$ and rotation with angle $\vec{\theta}$ we have respectively \cite{PhysRevD.93.084031}
\begin{align}
&L=B(\vec{v})=e^{\eta\hat{v}\cdot\vec{\sigma}},e^\eta=\sqrt{\gamma(1-v)},\gamma=\frac{1}{\sqrt{1-v^2}},\label{eq23}\\
&L=R(\vec{\theta})=e^{\frac{i}{2}\vec{\theta}\cdot\vec{\sigma}},
\end{align}
where $\vec{\sigma}=(\sigma^1,\sigma^2,\sigma^3)$ and $\sigma^i,i=1,2,3$ are the Pauli matrixes ((1.2.24) of \cite{PenRin88}). For boost $B(\vec{v})$ we have explicitly
\begin{align}
&B(\vec{v})=\begin{pmatrix}\cosh\eta+\frac{v_3}{v}\sinh\eta&(\frac{v_1}{v}+i\frac{v_2}{v})\sinh\eta\\
(\frac{v_1}{v}-i\frac{v_2}{v})\sinh\eta&\cosh\eta-\frac{v_3}{v}\sinh\eta\end{pmatrix},\label{eqII1}
\end{align}
where $\vec{v}=(v_1,v_2,v_3)$.

In the asymptotic region, the relation between Bondi-Sachs (BS) coordinate $(u,r,\theta,\phi)$ \cite{he2015new,he2016asymptotical,sun2019binary} and the above inertial Cartesian coordinate $(t,x,y,z)$ can be expressed as
\begin{align}
&t=u+r,x=r\sin\theta\cos\phi,y=r\sin\theta\sin\phi,z=r\cos\theta.
\end{align}
Correspondingly we can express the position matrix in (\ref{eqII2}) as
\begin{align}
\begin{pmatrix}t+z&x+iy\\x-iy&t-z\end{pmatrix}&=\begin{pmatrix}u+\frac{2r|\zeta|^2}{|\zeta|^2+1}&\frac{2r\zeta}{|\zeta|^2+1}\\\frac{2r\bar{\zeta}}{|\zeta|^2+1}&u+\frac{2r}{|\zeta|^2+1}\end{pmatrix},\\
\zeta&\equiv e^{i\phi}\cot\frac{\theta}{2},
\end{align}
where $\bar{\zeta}$ means the complex conjugate of $\zeta$.

Considering general transformation matrix
\begin{align}
L=\begin{pmatrix}a&b\\c&d\end{pmatrix}
\end{align}
which is hermitian, we have asymptotic BS coordinate transformation up to $O(\frac{1}{r})$
\begin{align}
u'&=ku,k\equiv\frac{1+\zeta\bar{\zeta}}{|a\zeta+b|^2+|c\zeta+d|^2},\label{eq24}\\
r'&=kr+\nonumber\\
&\frac{ku}{1+\zeta\bar{\zeta}}\left[2(a\bar{c}+b\bar{d})(c\zeta+d)(\bar{a}\bar{\zeta}+\bar{b})+\nonumber\right.\\
&2(\bar{a}c+\bar{b}d)(\bar{c}\bar{\zeta}+\bar{d})(a\zeta+b)+\nonumber\\
&\left.(|a|^2+|b|^2-|c|^2-|d|^2)(|a\zeta+b|^2-|c\zeta+d|^2)\right],\\
\zeta'&=\frac{a\zeta+b}{c\zeta+d}.
\end{align}
Here prime means the new BS coordinate.

We are free to choose the direction of BS coordinate. In order to simplify the calculation we let the $z$ axis point to the direction of the relative velocity. And more we choose the $x$ axis to let the gravitational wave source locate in the $x-z$ plane. Then $y$ axis is determined by the right-hand screw rule. Based on this choice of coordinate basis the source locates in the direction
\begin{align}
\theta\neq0,\phi=0,\label{eqII3}
\end{align}
and the Lorentz transformation matrix (\ref{eqII1}) can be simplified as
\begin{align}
B&=\begin{pmatrix}\cosh\eta+\sinh\eta&0\\
0&\cosh\eta-\sinh\eta\end{pmatrix}\\
&=\begin{pmatrix}e^\eta&0\\
0&e^{-\eta}\end{pmatrix},\label{eqII5}
\end{align}
which means
\begin{align}
a=e^\eta,d=e^{-\eta},b=c=0.
\end{align}
So the above general transformation becomes
\begin{align}
u'&=ku,k\equiv\frac{1+|\zeta|^2}{a^2|\zeta|^2+d^2},\\
r'&=\frac{r}{k}+\frac{u}{k}\frac{(|a|^2-|d|^2)(a^2|\zeta|^2-d^2)}{1+|\zeta|^2},\\
\zeta'&=\frac{a}{d}\zeta,\label{eqII4}
\end{align}
\section{Lorentz transformation of electromagnetic wave within the BMS theory}
In the asymptotic region, or to say the wave zone, the BS coordinate basis $\hat{r}$ corresponds to the propagating direction of electromagnetic (EM) wave. Based on the property of EM wave we have $\hat{r}\cdot\vec{E}=0,\vec{B}=\hat{r}\times\vec{E},\vec{E}=\vec{B}\times\hat{r}$. Using the tetrad $(\hat{t},\hat{r},\hat{\theta},\hat{\phi})$ we have EM tensor field $F_{\mu\nu}$ and the Newman-Penrose tetrad as following
\begin{align}
&F_{\mu\nu}=\begin{pmatrix}0&E_{\hat{\theta}}&E_{\hat{\phi}}&0\\
                        -E_{\hat{\theta}}&0&0&E_{\hat{\theta}}\\
                        -E_{\hat{\phi}}&0&0&E_{\hat{\phi}}\\
                         0  &-E_{\hat{\theta}}&-E_{\hat{\phi}}&0
                         \end{pmatrix},\\
&l^a=\frac{1}{\sqrt{2}}(\hat{t}^a+\hat{r}^a),n^a=\frac{1}{\sqrt{2}}(\hat{t}^a-\hat{r}^a),m^a=\frac{1}{\sqrt{2}}(\hat{\theta}^a+i\hat{\phi}).
\end{align}
Then we have Newman-Penrose EM scalar
\begin{align}
&\phi_2\equiv F_{ab}n^a\bar{m}^b=E_{\hat{\theta}}-iE_{\hat{\phi}}.\label{eq8}
\end{align}

The boost BMS transformation results in \cite{PenRin88}
\begin{align}
&\phi_2'=\frac{e^{-i\lambda}}{k}\phi_2,\label{eqII7}\\
&e^{i\lambda}=\frac{c\zeta+d}{\overline{c\zeta+d}}.\label{eqII6}
\end{align}
And the EM propagating direction will change from $\hat{r}$ to $\hat{r}'$. Again here the prime means the new coordinate and the new frame after the Lorentz transformation. Specifically the direction is described by $\theta$ and $\theta'$ due to the property (\ref{eqII3}). Together with (\ref{eqII5}) and (\ref{eq23}) the transformation (\ref{eqII4}) results in
\begin{align}
&\cot\frac{\theta'}{2}=\gamma(1-v)\cot\frac{\theta}{2},\label{eqII12}
\end{align}
which is nothing but the usual aberration formula \cite{1998clel.book.....J}. The above aberration formula (\ref{eqII12}) can also be expressed as
\begin{align}
&\cos\theta'=\frac{\cos\theta-v}{1-v\cos\theta},\sin\theta'=\frac{\sin\theta}{\gamma(1-v\cos\theta)}.\label{eqII14}
\end{align}

From (\ref{eqII7}) we can see a phase change $e^{-i\lambda}$ which corresponds to `spin weight 1', and an amplitude change $\frac{1}{k}$  which corresponds to `boost weight 1'. Altogether we conclude that EM wave behaves as a rank-one tensor which is consistent to usual understanding that EM wave is a vector field.

Due to (\ref{eqII5}), (\ref{eqII6}) becomes
\begin{align}
&e^{i\lambda}=1,\lambda=0\label{eq19},\\
&k=\frac{1}{\gamma(1-\vec{v}\cdot\hat{r})}.
\end{align}
Consequently (\ref{eqII7}) results in
\begin{align}
&\phi_2'=\frac{1}{k}\phi_2,\\
&E'_{\hat{\theta}}=\frac{E_{\hat{\theta}}}{k},E'_{\hat{\phi}}=\frac{E_{\hat{\phi}}}{k}.\label{eqII13}
\end{align}
Noting the relation between the Cartesian frame and the spherical frame
\begin{align}
&\hat{r}=\sin\theta\hat{e}_x+\cos\theta\hat{e}_z,\\
&\hat{\theta}=\cos\theta\hat{e}_x-\sin\theta\hat{e}_z,\\
&\hat{\phi}=\hat{e}_y,
\end{align}
if we use three dimensional vector to express the electric field, we have
\begin{align}
\vec{E}&=E_\theta\hat{e}_\theta+E_\phi\hat{e}_\phi,\\
&=E_\theta\cos\theta\hat{e}_x+E_\phi\hat{e}_y-E_\theta\sin\theta\hat{e}_z,\\
\vec{E}'&=E'_\theta\cos\theta'\hat{e}'_x+E'_\phi\hat{e}'_y-E'_\theta\sin\theta'\hat{e}'_z.\label{eqII11}
\end{align}
Frame $(x,y,z)$ deems frame $(x',y',z')$ moves in $z$ direction, while frame $(x',y',z')$ deems frame $(x,y,z)$ moves in $-z'$ direction. Both frames agree that the relative velocity lies in the same line. Or to say they deem $\hat{e}_z$ and $\hat{e}'_z$ point to the same direction. In addition since both $\hat{e}_z$ and $\hat{e}'_z$ admit unit length we have
\begin{align}
\hat{e}_z=\hat{e}'_z.\label{eqII9}
\end{align}
According to the Lorentz transformation between $(t,x,y,z)$ and $(t',x',y',z')$
\begin{align}
t'&=\gamma (t-vz),\\
x'&=x,\\
y'&=y,\\
z'&=\gamma (z-vt),
\end{align}
we straightforwardly have
\begin{align}
\hat{e}_x=\hat{e}'_x,\, \hat{e}_y=\hat{e}'_y.\label{eqII10}
\end{align}
Plugging the relations (\ref{eqII9}) and (\ref{eqII10}) into (\ref{eqII11}) we get
\begin{align}
\vec{E}'&=E'_\theta\cos\theta'\hat{e}_x+E'_\phi\hat{e}_y-E'_\theta\sin\theta'\hat{e}_z.
\end{align}
Combining relations (\ref{eqII14}) and (\ref{eqII13}) we have
\begin{align}
\vec{E}'&=\gamma(1-v\cos\theta)\times\nonumber\\
&\left(E_\theta\frac{\cos\theta-v}{1-v\cos\theta}\hat{e}_x+E_\phi\hat{e}_y-E_\theta\frac{\sin\theta}{\gamma(1-v\cos\theta)}\hat{e}_z\right)\\
&=\gamma\vec{E}(1-v\cos\theta)\nonumber\\
&-vE_\theta\sin\theta\left(\gamma\cos\theta\hat{e}_z+\gamma\sin\theta\hat{e}_x-\frac{\gamma^2}{1+\gamma}v\hat{e}_z\right)\\
&=\gamma(1-\vec{v}\cdot\hat{r})\vec{E}+\gamma(\vec{v}\cdot\vec{E})(\hat{r}-\frac{\gamma}{1+\gamma}\vec{v}).\label{eqII15}
\end{align}
We can find that the above result is consistent to the usual Lorentz transformation of electromagnetic field \cite{1998clel.book.....J}
\begin{align}
\vec{E}'&=\gamma(\vec{E}+\vec{v}\times\vec{B})-\frac{\gamma^2}{1+\gamma}\vec{v}\cdot\vec{E}\vec{v}\\
&=\gamma(1-\vec{v}\cdot\hat{r})\vec{E}+\gamma(\vec{v}\cdot\vec{E})(\hat{r}-\frac{\gamma}{1+\gamma}\vec{v}).\label{eqII16}
\end{align}
In the last step we have used EM wave relation $\vec{B}=\hat{r}\times\vec{E}$. The consistence between (\ref{eqII15}) and (\ref{eqII16}) verifies relations (\ref{eqII9}) and (\ref{eqII10}) which will be used to deduce Lorentz transformation of gravitational wave in the next subsection.

\section{Lorentz transformation of gravitational wave}
Within the tetrad $(\hat{t},\hat{r},\hat{\theta},\hat{\phi})$ introduced in the last section, gravitational wave can be expressed as
\begin{align}
h_{ij}&=h_+e^+_{ij}+h_\times e^\times_{ij},\label{eq18}\\
e^+_{ij}&=\hat{\theta}_i\hat{\theta}_j-\hat{\phi}_i\hat{\phi}_j\nonumber\\
&=\cos^2\theta\hat{e}_x\hat{e}_x-\sin2\theta\hat{e}_x\hat{e}_z+\sin^2\theta\hat{e}_z\hat{e}_z-\hat{e}_y\hat{e}_y\label{eq29}\\
e^\times_{ij}&=\hat{\theta}_i\hat{\phi}_j+\hat{\theta}_j\hat{\phi}_i\nonumber\\
&=2\cos\theta\hat{e}_x\hat{e}_y-2\sin\theta\hat{e}_y\hat{e}_z,\label{eq30}
\end{align}
where $h_{+,\times}$ corresponds to the two polarization modes of gravitational wave.

On the other hand we can express the BMS transformation of gravitational wave with notation $h\equiv h_+-ih_{\times}$ as
\begin{align}
&h'=e^{-i2\lambda}(h-\bar{\eth}^2\alpha),\label{eq17}
\end{align}
where $\alpha$ corresponds to super-translation. If we only care about Lorentz transformation, $\alpha$ vanishes. Our above equation is different to $h'=\frac{e^{-i2\lambda}}{k}(h-\bar{\eth}^2\alpha)$ (Eq.~(21) of \cite{PhysRevD.93.084031}) at the first sight, but as pointed out in the footnote 5 of \cite{PhysRevD.93.084031}, our $h$ means the physical gravitational wave strain which corresponds to $H$ of \cite{PhysRevD.93.084031}.

From (\ref{eq17}) we can see a phase change $e^{-i2\lambda}$ which corresponds to `spin weight 2', while the amplitude keeps unchanged. In another word, the amplitude changes as $\frac{1}{k^0}$  which corresponds to `boost weight 0'. Unlike the EM wave, gravitational wave admits different weight factor for spin and boost. Or to say GW behaves as either tensor field or scalar field. But GW admit both characters of tensor field and scalar field. Just because GW partially behaves like a scalar field, ones can treat GW lensing as a scalar wave \cite{RyuTak03,PhysRevD.90.062003,PhysRevD.98.104029}. Just because GW partially behaves like a tensor field, people project the GW `tensor' (\ref{eq18}) onto a detector when considering the response of a detector to a given GW \cite{maggiore2008gravitational}.

Based on the BMS transformation for Lorentz transformation $\alpha=0$, and the setting in the previous section which results in (\ref{eq19}), we have
\begin{align}
&h'=h,\\
&h_+'=h_+,h_{\times}'=h_{\times}.\label{eq28}
\end{align}
Consequently the GW `tensor' after Lorentz transformation reads as
\begin{align}
h'_{ij}&=h'_+e'^+_{ij}+h'_\times e'^\times_{ij}\label{eq21}\\
&=h_+\left(\cos^2\theta'\hat{e}_x\hat{e}_x-\sin2\theta'\hat{e}_x\hat{e}_z+\sin^2\theta'\hat{e}_z\hat{e}_z-\hat{e}_y\hat{e}_y\right)\nonumber\\
&+h_\times
\left(2\cos\theta'\hat{e}_x\hat{e}_y-2\sin\theta'\hat{e}_y\hat{e}_z\right),
\end{align}
where the relations (\ref{eqII9}) and (\ref{eqII10}) have been used. Finally plugging the aberration formula (\ref{eqII14}) into the above equation we get
\begin{align}
h'_{ij}&=h_+\left((\frac{\cos\theta-v}{1-v\cos\theta})^2\hat{e}_x\hat{e}_x
-2\frac{(\cos\theta-v)\sin\theta}{\gamma(1-v\cos\theta)^2}\hat{e}_x\hat{e}_z\right.\nonumber\\
&\left.+(\frac{\sin\theta}{\gamma(1-v\cos\theta)})^2\hat{e}_z\hat{e}_z-\hat{e}_y\hat{e}_y\right)\nonumber\\
&+h_\times
\left(2\frac{\cos\theta-v}{1-v\cos\theta}\hat{e}_x\hat{e}_y-2\frac{\sin\theta}{\gamma(1-v\cos\theta)}\hat{e}_y\hat{e}_z\right).
\end{align}
Equivalently we can express the above transformation as a tensor form
\begin{align}
h'_{ij}&=h_{ij}+v^kh_{kl}v^l\frac{1}{(1-\hat{r}\cdot\vec{v})^2}\left[\hat{r}_i\hat{r}_j\right.\nonumber\\
&\quad\quad\quad\quad\quad\left.-\frac{\gamma}{1+\gamma}(\hat{r}_iv_j+v_i\hat{r}_j)+\frac{\gamma^2}{(1+\gamma)^2}v_iv_j\right]\nonumber\\
&+v^kh_{kj}\frac{1}{1-\hat{r}\cdot\vec{v}}[\hat{r}_i-\frac{\gamma}{1+\gamma}v_i]\nonumber\\
&+v^kh_{ik}\frac{1}{1-\hat{r}\cdot\vec{v}}[\hat{r}_j-\frac{\gamma}{1+\gamma}v_j],\label{eq20}
\end{align}
which is independent of coordinate choice. This is the Lorentz transformation formula of GW from a rest frame to a moving frame with velocity $\vec{v}$. In this form the velocity is not limited along $z$ direction. In stead it can point to any direction.

At the first glance, the transformation may diverge when the relative velocity is along the direction of the GW propagates and the relative velocity tends to the speed of light $\hat{r}\cdot\vec{v}\rightarrow1$. In such case $v^kh_{kl}=0$ due to the transverse property of GW. Consequently $h'_{ij}=h_{ij}$ when the relative velocity is along the direction of the GW propagates.

As shown in (1.12) of \cite{maggiore2008gravitational}, GW behaves as a spin-2 tensor if only the Lorentz transformation velocity is small which guarantees $|h_{\mu\nu}|\ll1$. Up to the first order of the relative velocity $v$, the Lorentz transformation matrix (corresponding to $\Lambda^\mu{}_\nu$ in (1.9) of \cite{maggiore2008gravitational}) takes form
\begin{align}
\Lambda^\mu{}_\nu=\begin{pmatrix}\gamma&-\gamma v_i\\-\gamma v_j&\delta_{ij}-\frac{1-\gamma}{v^2}v_iv_j\end{pmatrix}\approx\begin{pmatrix}1&-v_i\\-v_j&\delta_{ij}\end{pmatrix}.
\end{align}
According to the transformation (1.12) of \cite{maggiore2008gravitational} and paying additional attention to the transverse-traceless gauge transformation, we will get
\begin{align}
h'_{ij}&=h_{ij}+v^kh_{kj}\hat{r}_i+v^kh_{ik}\hat{r}_j,
\end{align}
which is consistent to our Lorentz transformation formula (\ref{eq20}). In order to deduce the above result we have used the propagating wave property of $h_{ij}$ which requires the dependence of $h_{ij}$ on space and time just through $(t-\hat{r}\cdot\vec{x})$. Here $\vec{x}$ denotes the position vector.

It can be checked straight forwardly that $h'_{ij}$ in (\ref{eq20}) is traceless and $h'_{ij}\hat{r}'^i=0$ which means $h'_{ij}$ is transverse. This is to say our Lorentz transformation preserves the transverse-traceless property of gravitational wave tensor.

In addition we can note that $h_{ij}h^{ij}=h^2_++h^2_\times$. The relation (\ref{eq17}) indicates that the Lorentz transformation admits $h'=he^{-2\lambda}$ and consequently
\begin{align}
h^2_++h^2_\times=h'^2_++h'^2_\times.
\end{align}
As a self consistent check, ones can show that the Lorentz transformation formula (\ref{eq20}) does result in $h_{ij}h^{ij}=h'_{ij}h'^{ij}$. The calculation is straightforward but tedious. A trick is denoting the $h'_{ij}$ in (\ref{eq20}) as
\begin{align}
&h'_{ij}=h_{ij}+p_{ij}+q_{ij}+s_{ij},\\
&p_{ij}=v^kh_{kl}v^l\frac{1}{(1-\hat{r}\cdot\vec{v})^2}\left[\hat{r}_i\hat{r}_j\right.\nonumber\\
&\quad\quad\quad\quad\quad\left.-\frac{\gamma}{1+\gamma}(\hat{r}_iv_j+v_i\hat{r}_j)+\frac{\gamma^2}{(1+\gamma)^2}v_iv_j\right],\\
&q_{ij}=v^kh_{kj}\frac{1}{1-\hat{r}\cdot\vec{v}}[\hat{r}_i-\frac{\gamma}{1+\gamma}v_i],\\
&s_{ij}=v^kh_{ik}\frac{1}{1-\hat{r}\cdot\vec{v}}[\hat{r}_j-\frac{\gamma}{1+\gamma}v_j].
\end{align}
Then we have
\begin{align}
h'_{ij}h'^{ij}=h_{ij}h^{ij}&+p_{ij}p^{ij}+2q_{ij}q^{ij}+2h_{ij}p^{ij}\nonumber\\
&+4h_{ij}q^{ij}+4p_{ij}q^{ij}+2q_{ij}s^{ij}.
\end{align}
Here we have used property $q_{ij}=s_{ji}$.

Using relation
\begin{align}
1+\frac{\gamma^2v^2}{(1+\gamma)^2}=\frac{2\gamma}{1+\gamma}.\label{eqIII31}
\end{align}
we can get
\begin{align}
q_{ij}q^{ij}=2h_{ij}q^{ij}.
\end{align}

Repeatedly using the relation (\ref{eqIII31}) we can get
\begin{align}
p_{ij}p^{ij}+2h_{ij}p^{ij}+4p_{ij}q^{ij}+2q_{ij}s^{ij}=0,
\end{align}
which results in
\begin{align}
h'_{ij}h'^{ij}=h_{ij}h^{ij}.
\end{align}

For those readers who take the gravitational wave as a perturbation of flat spacetime, gravitational wave can be described as a four dimensional tensor. They may be interested in how about the Lorentz transformation of such a four dimensional tensor. Actually such transformation can be got quite easily. The three dimensional tensor discussed above exactly corresponds to the spacial part of such a four dimensional tensor. And due to the transverse-traceless requirement, the time related components all vanishes. So just complementing one row and one column zeros to the three dimensional tensor got by our Lorentz transformation rule, ones can get the four dimensional GW tensor.
\section{Calculation of phase change due to boost for general velocity}
\begin{figure}
\begin{tabular}{c}
\includegraphics[width=0.48\textwidth]{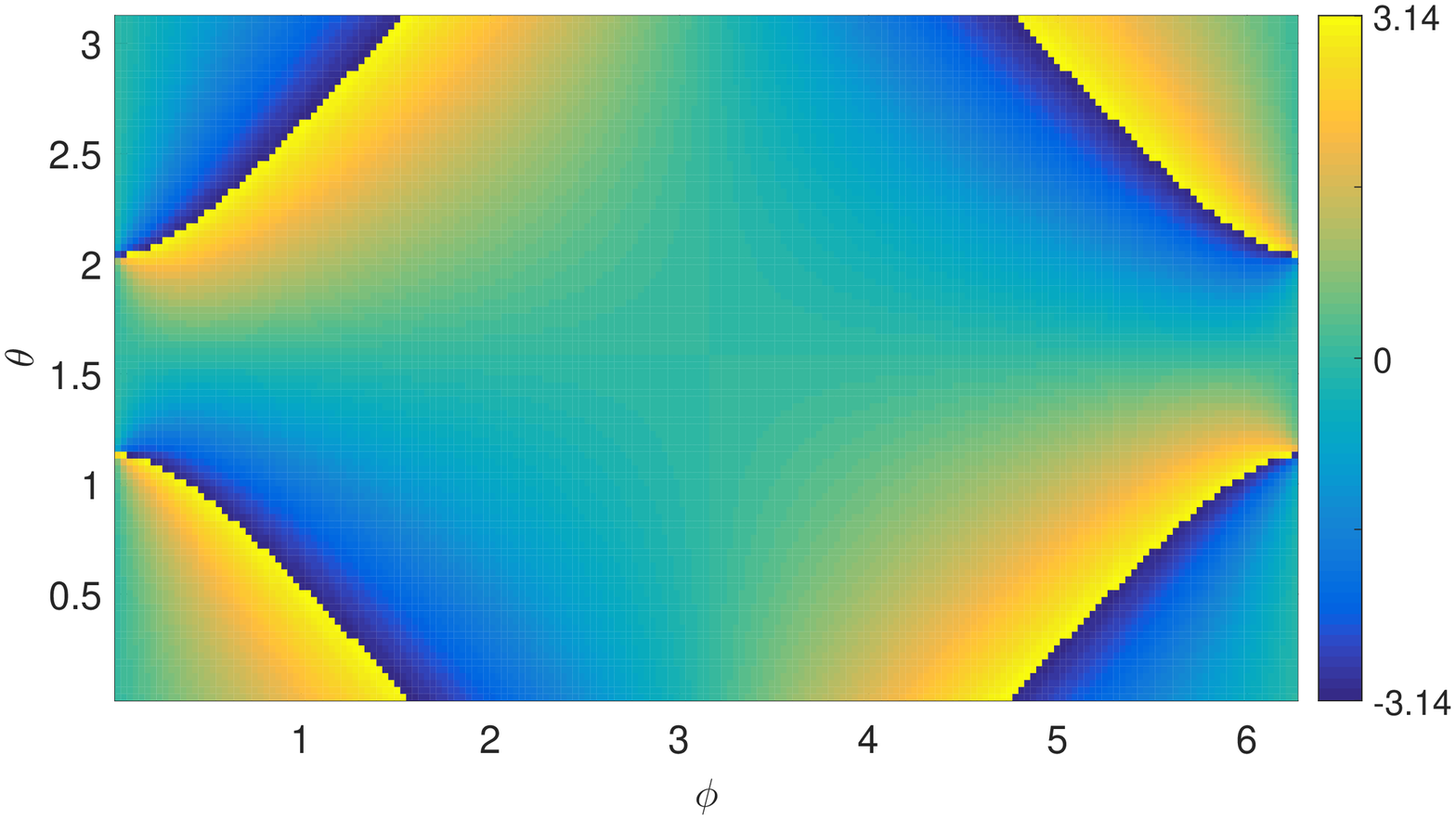}\\
\includegraphics[width=0.48\textwidth]{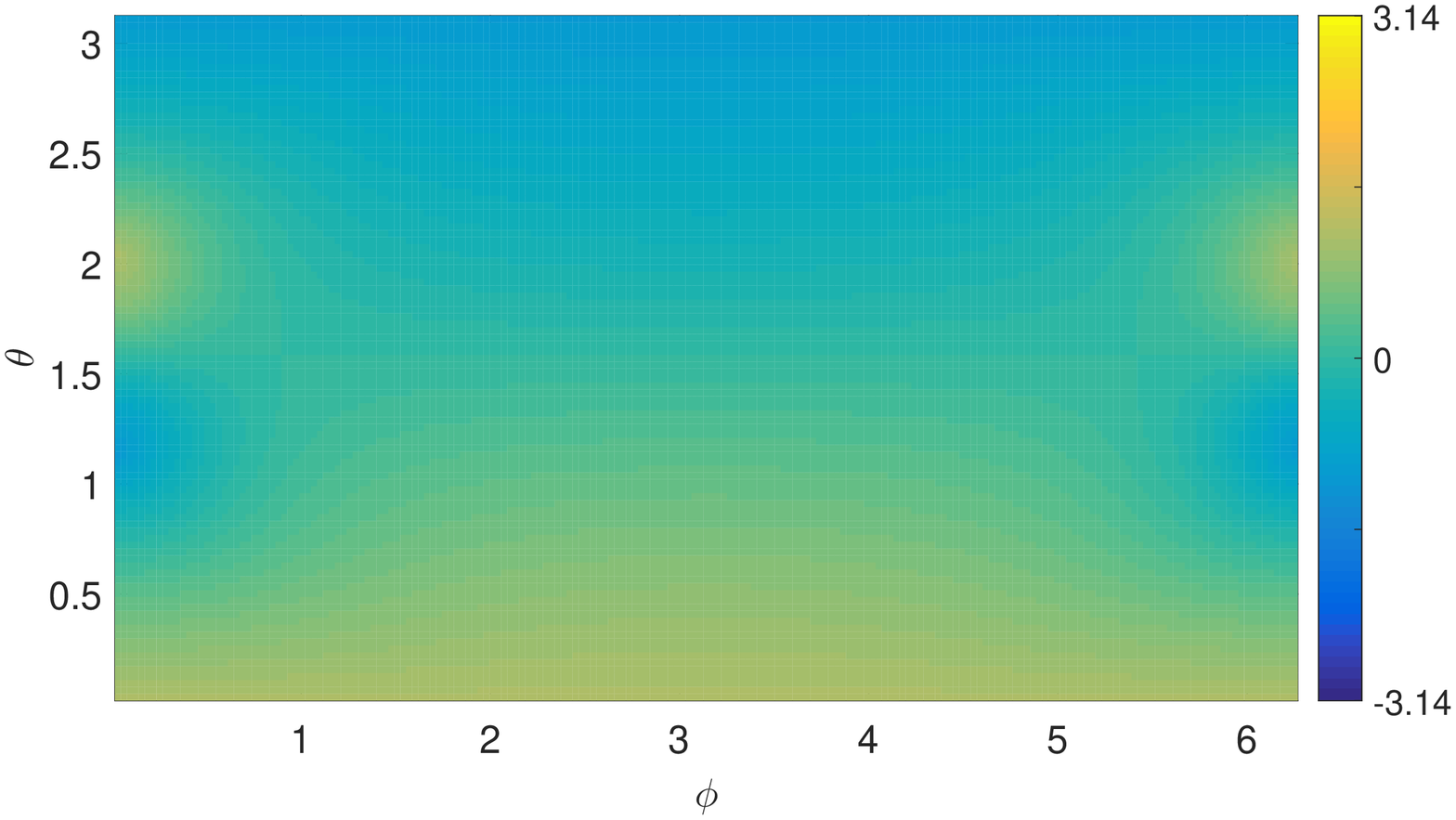}\\
\includegraphics[width=0.48\textwidth]{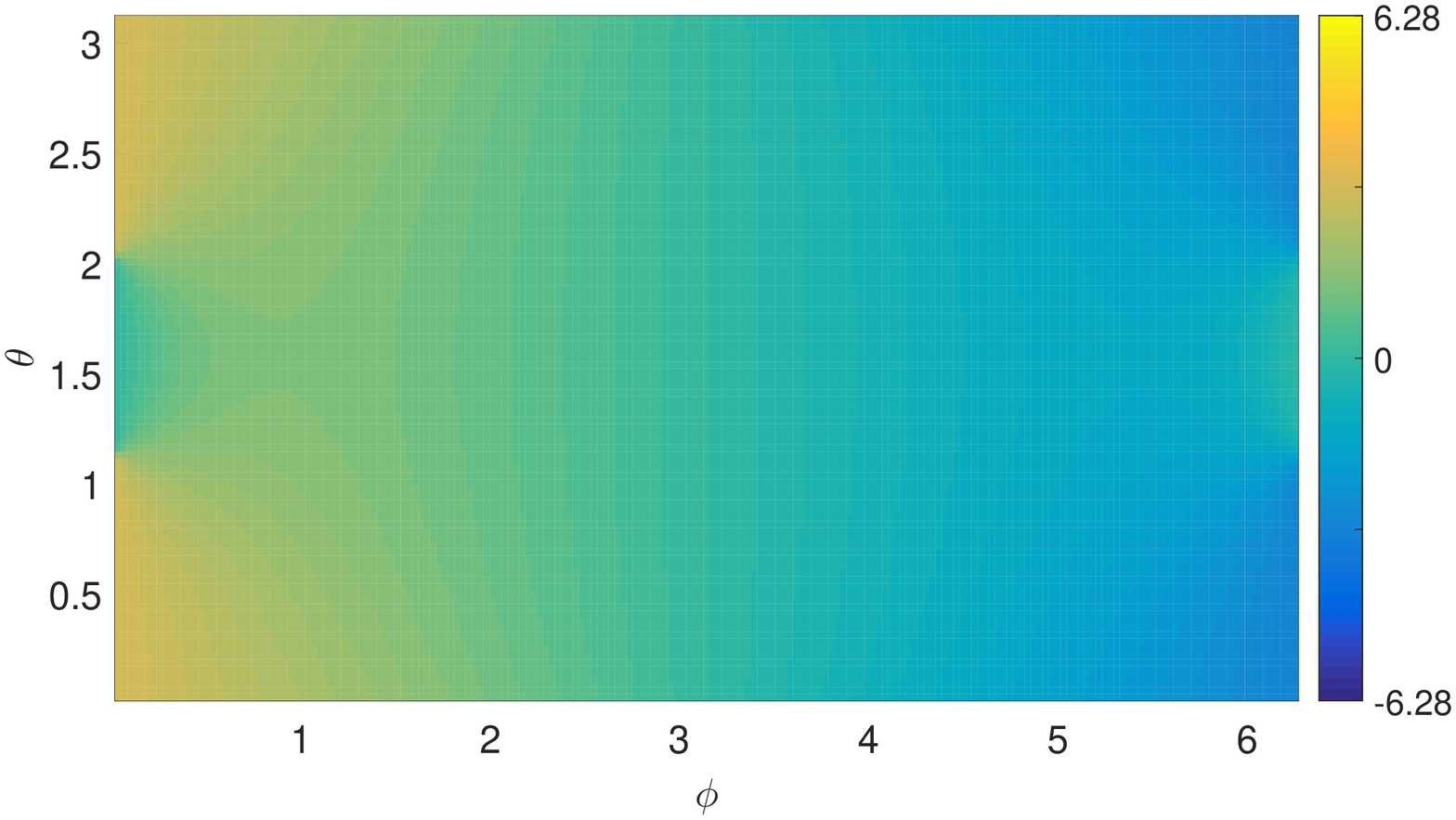}
\end{tabular}
\caption{Phase change and aberration angle for boost velocity $\vec{v}=(0.9,0,0)$. The top panel is for the phase change $\lambda\in(-\pi,\pi)$. The middle panel is for $\Delta \theta\equiv\theta'-\theta$. And the bottom panel is for $\Delta \phi\equiv\phi'-\phi$. The phase change seems to admit unsmooth jumps. That is because $-\pi$ and $\pi$ should be identified but the plot shows a jump from $-\pi$ to $\pi$.}\label{fig1}
\end{figure}
\begin{figure}
\begin{tabular}{c}
\includegraphics[width=0.48\textwidth]{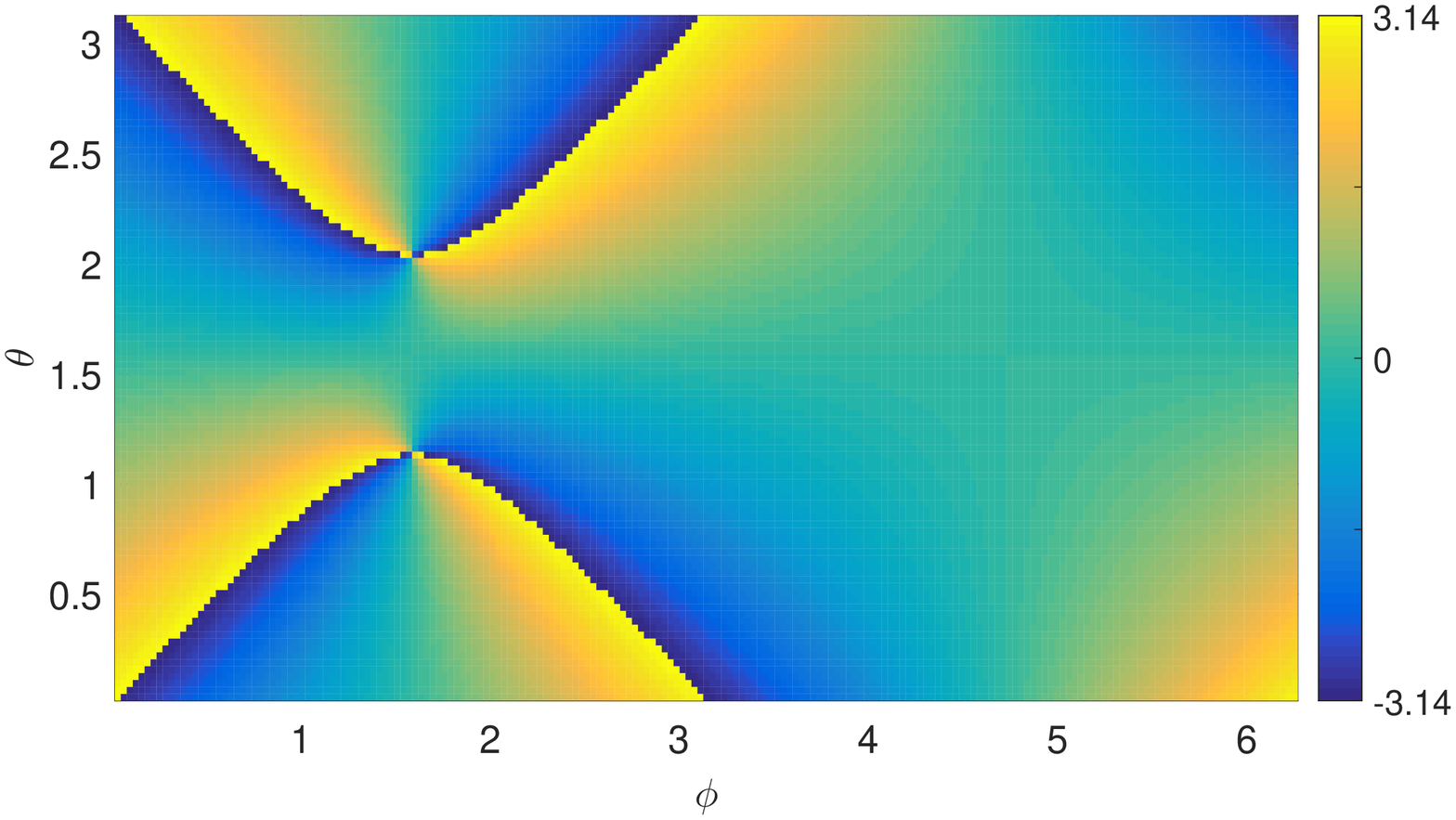}\\
\includegraphics[width=0.48\textwidth]{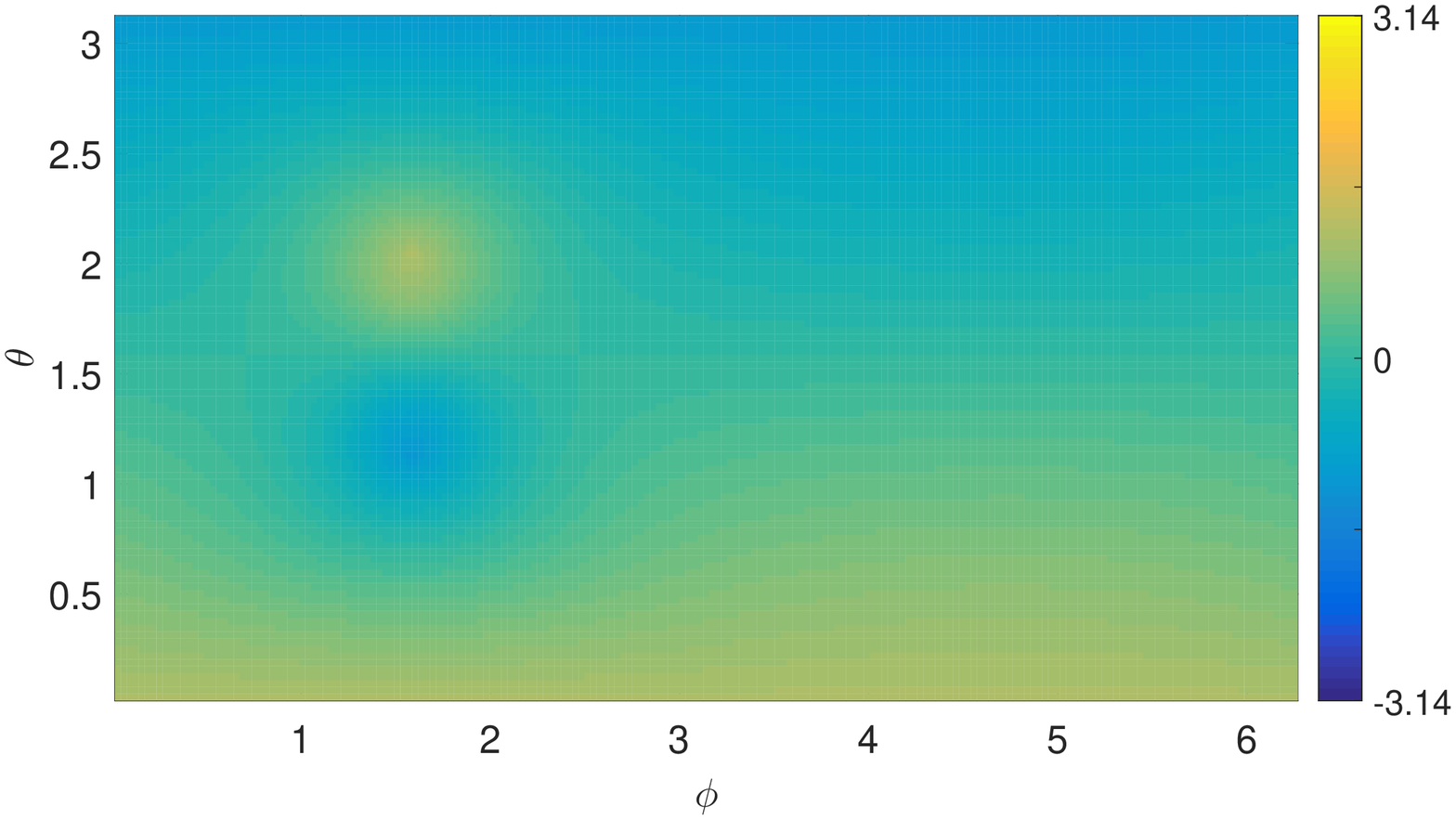}\\
\includegraphics[width=0.48\textwidth]{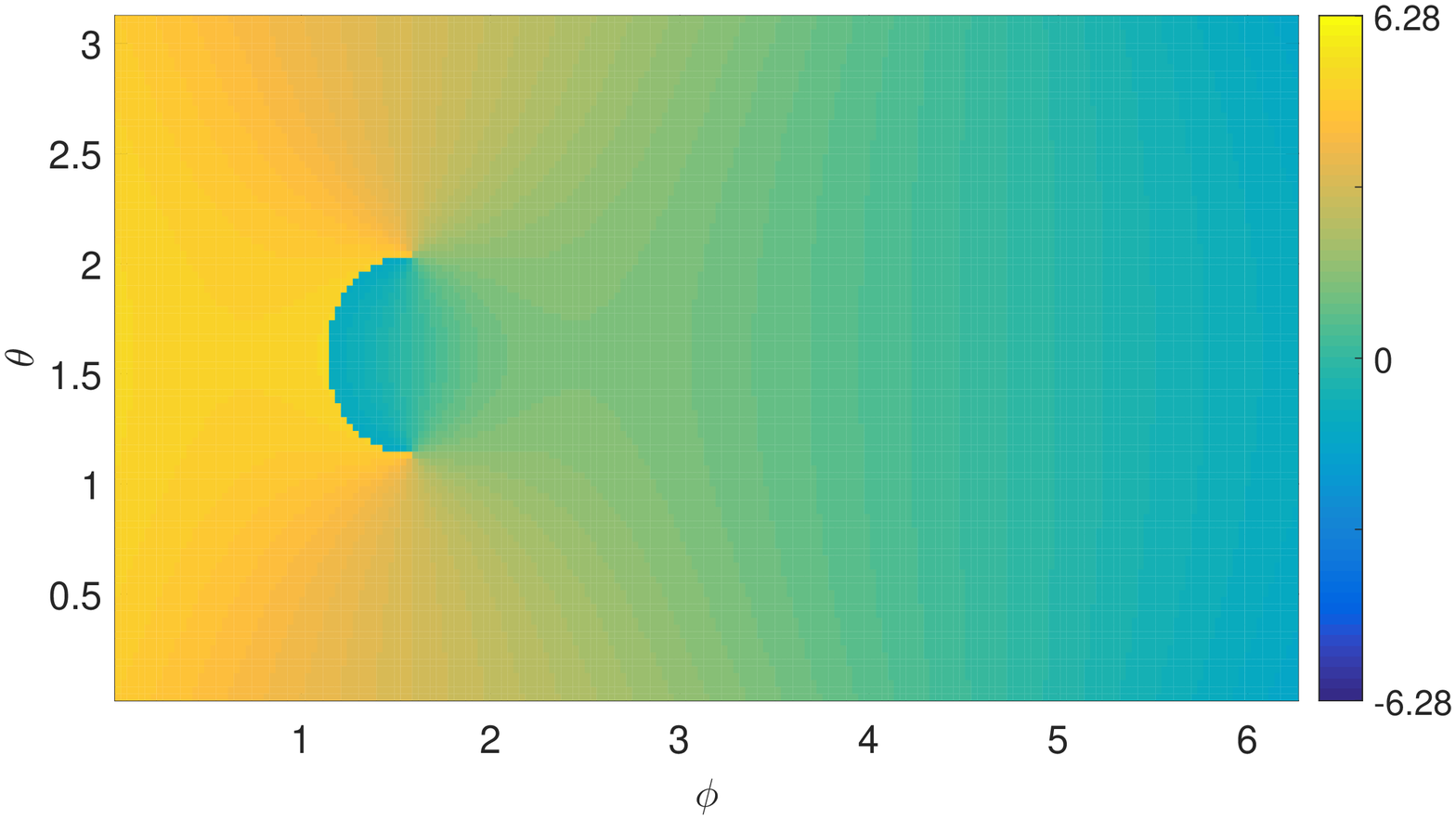}
\end{tabular}
\caption{The plot convention is the same to Fig.~\ref{fig1} but here the boost velocity is $\vec{v}=(0,0.9,0)$. The bottom panel seems to admit unsmooth jumps. This is because $\phi'$ takes values in $(0,2\pi)$ while 0 and $2\pi$ should be identified. The apparent jump corresponds to the contact place of $\phi'=0$ and $\phi'=2\pi$.}\label{fig2}
\end{figure}
\begin{figure}
\begin{tabular}{c}
\includegraphics[width=0.48\textwidth]{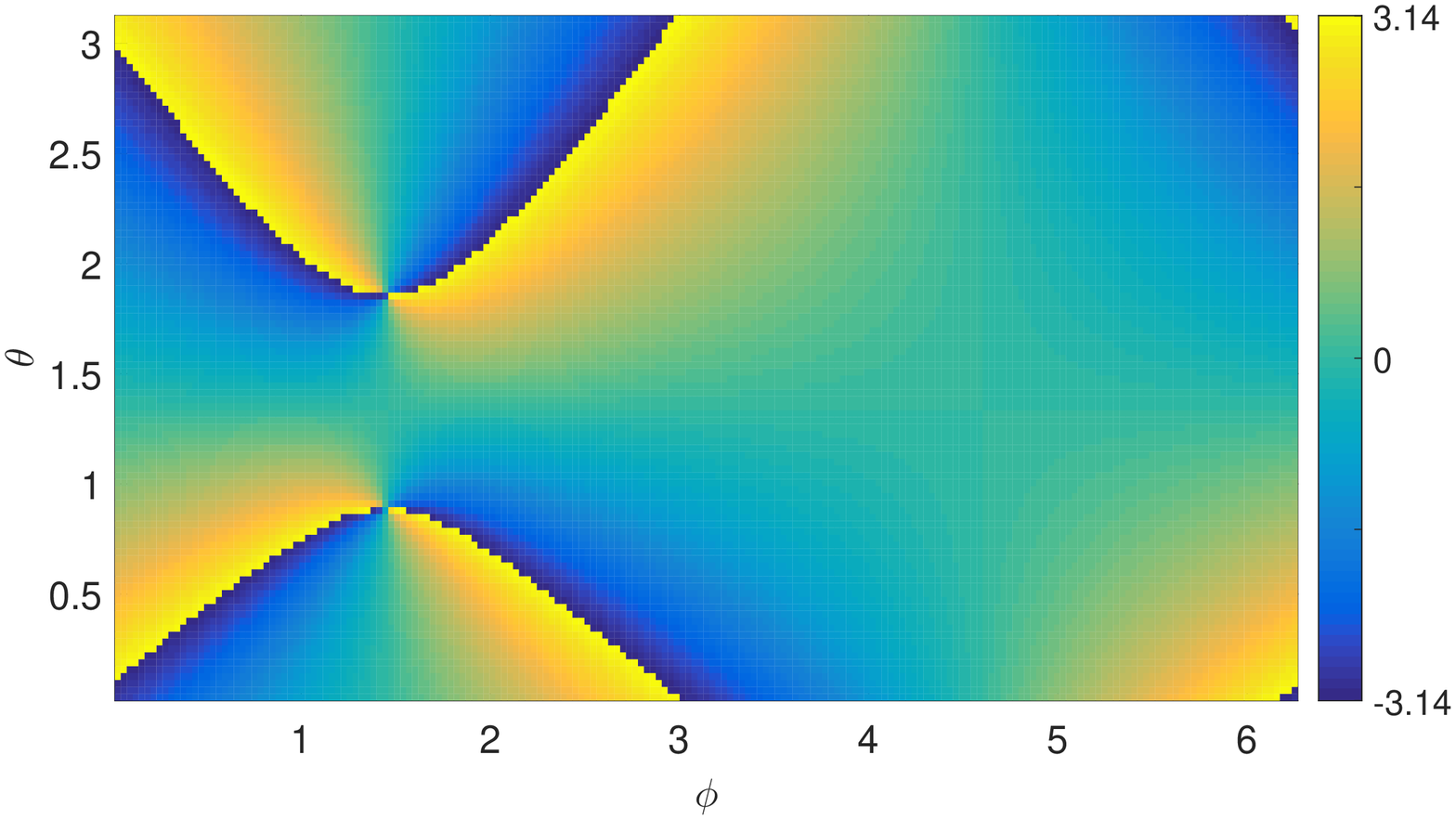}\\
\includegraphics[width=0.48\textwidth]{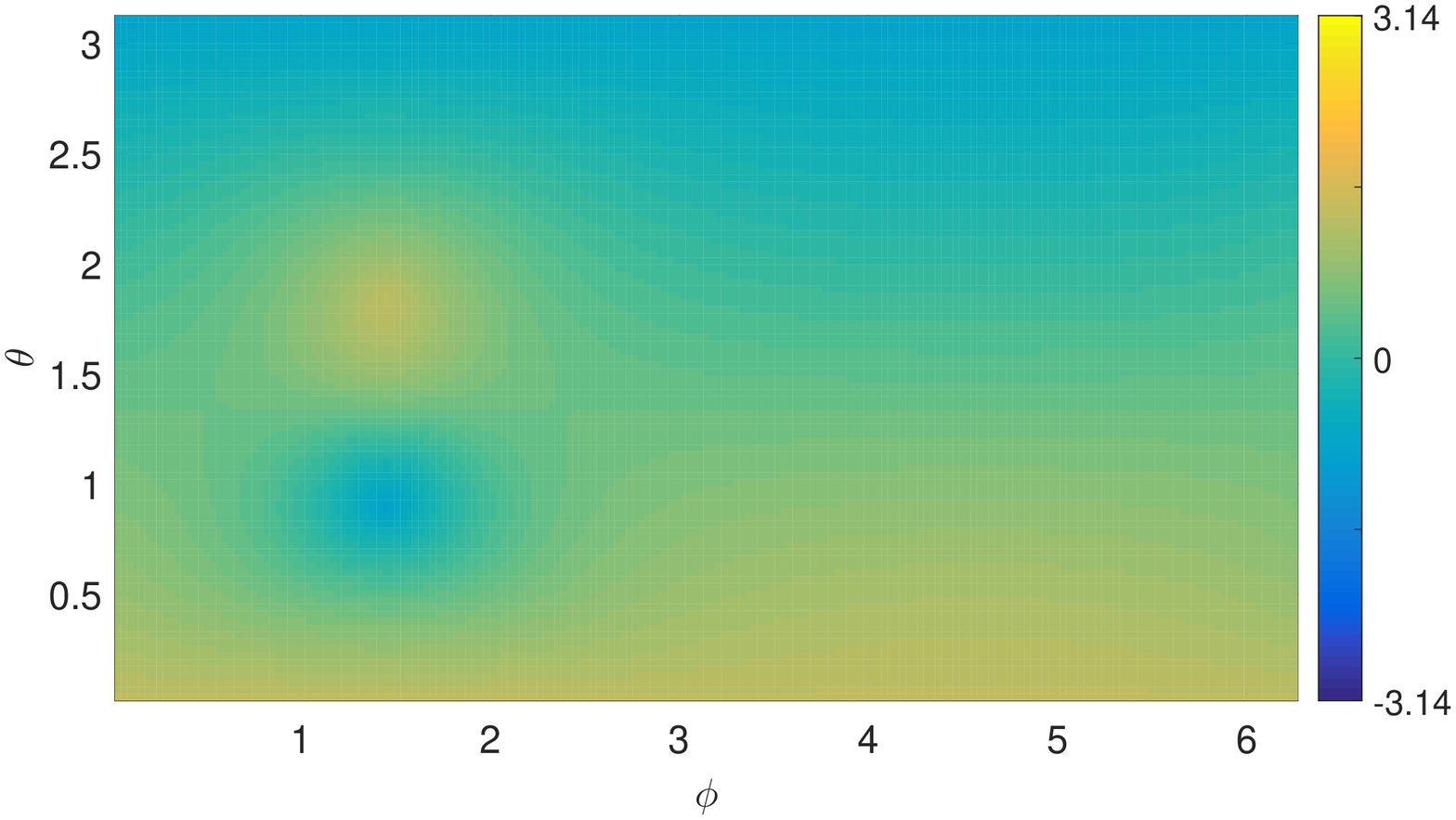}\\
\includegraphics[width=0.48\textwidth]{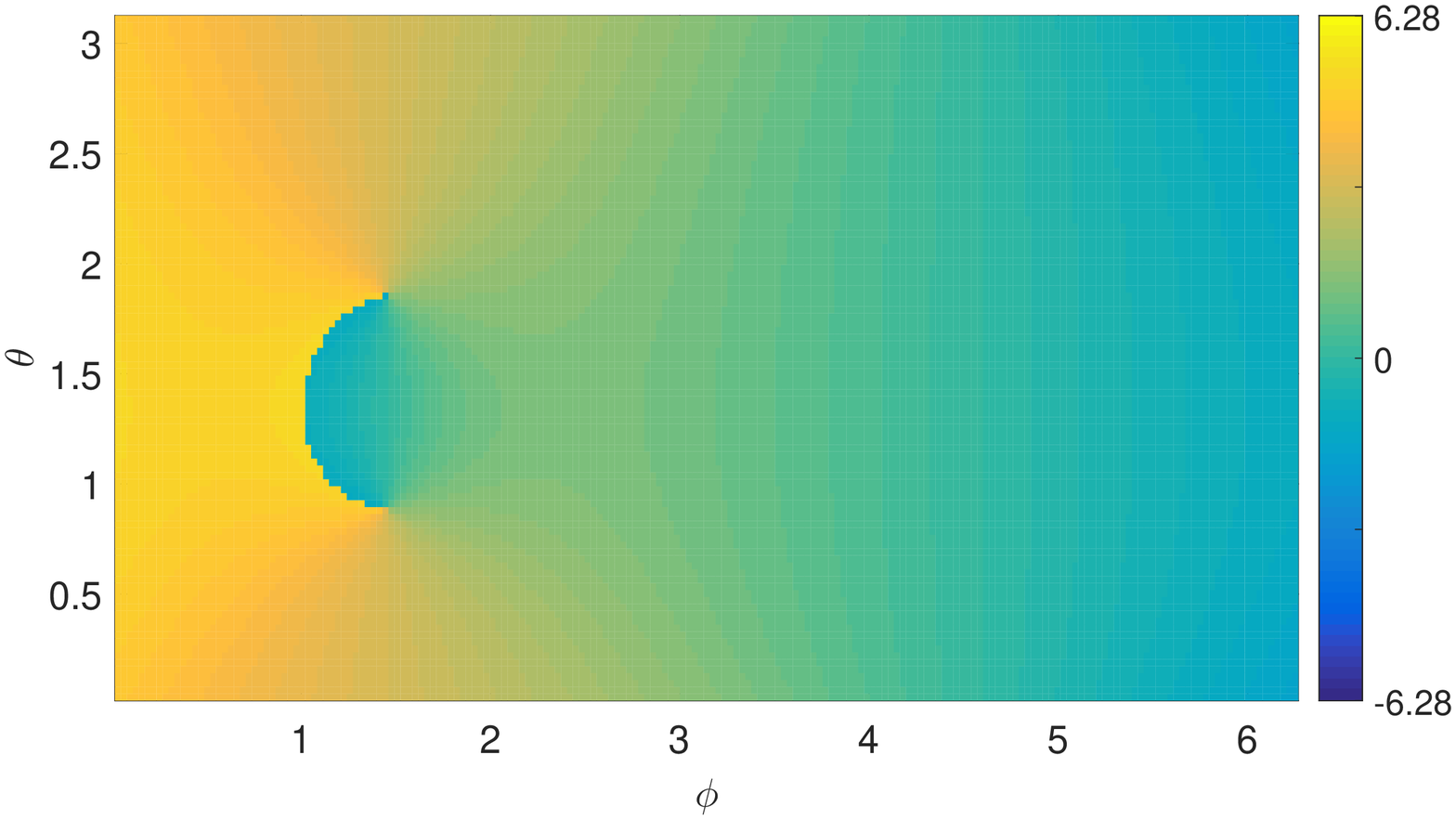}
\end{tabular}
\caption{The plot convention is the same to Fig.~\ref{fig1} but here the boost velocity is $\vec{v}=(0.1143,0.8219,0.3481)$.}\label{fig3}
\end{figure}
\begin{figure}
\begin{tabular}{c}
\includegraphics[width=0.48\textwidth]{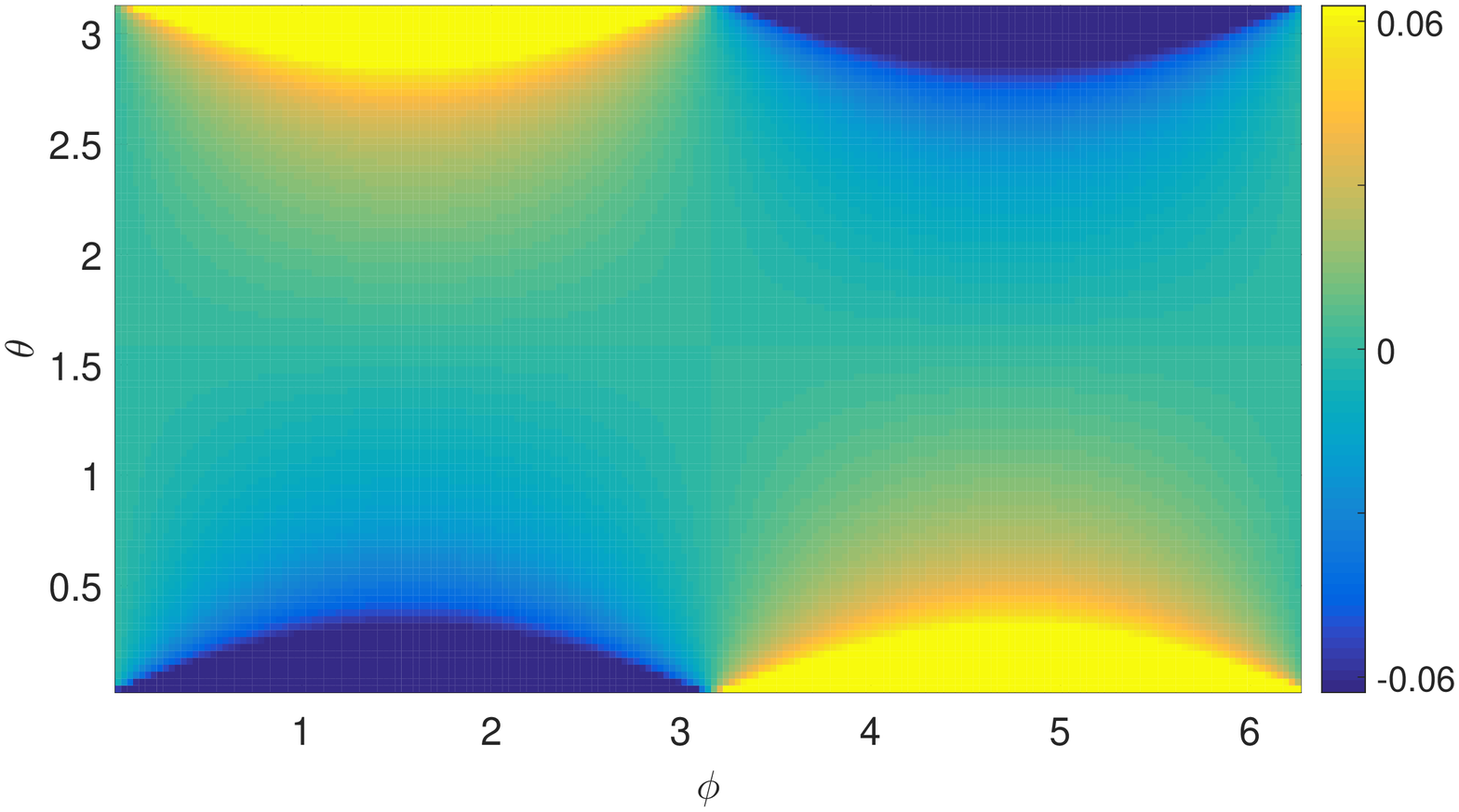}\\
\includegraphics[width=0.48\textwidth]{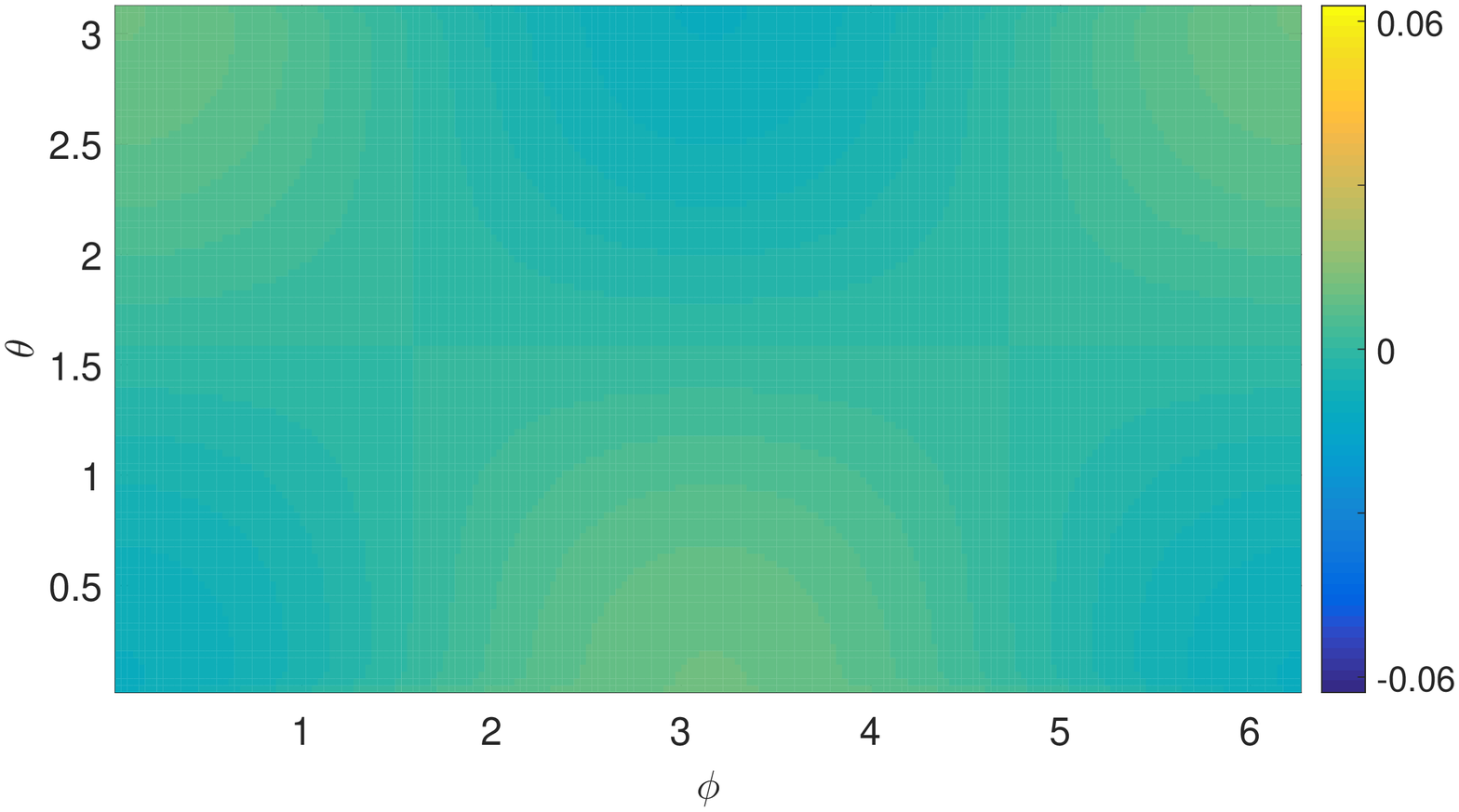}\\
\includegraphics[width=0.48\textwidth]{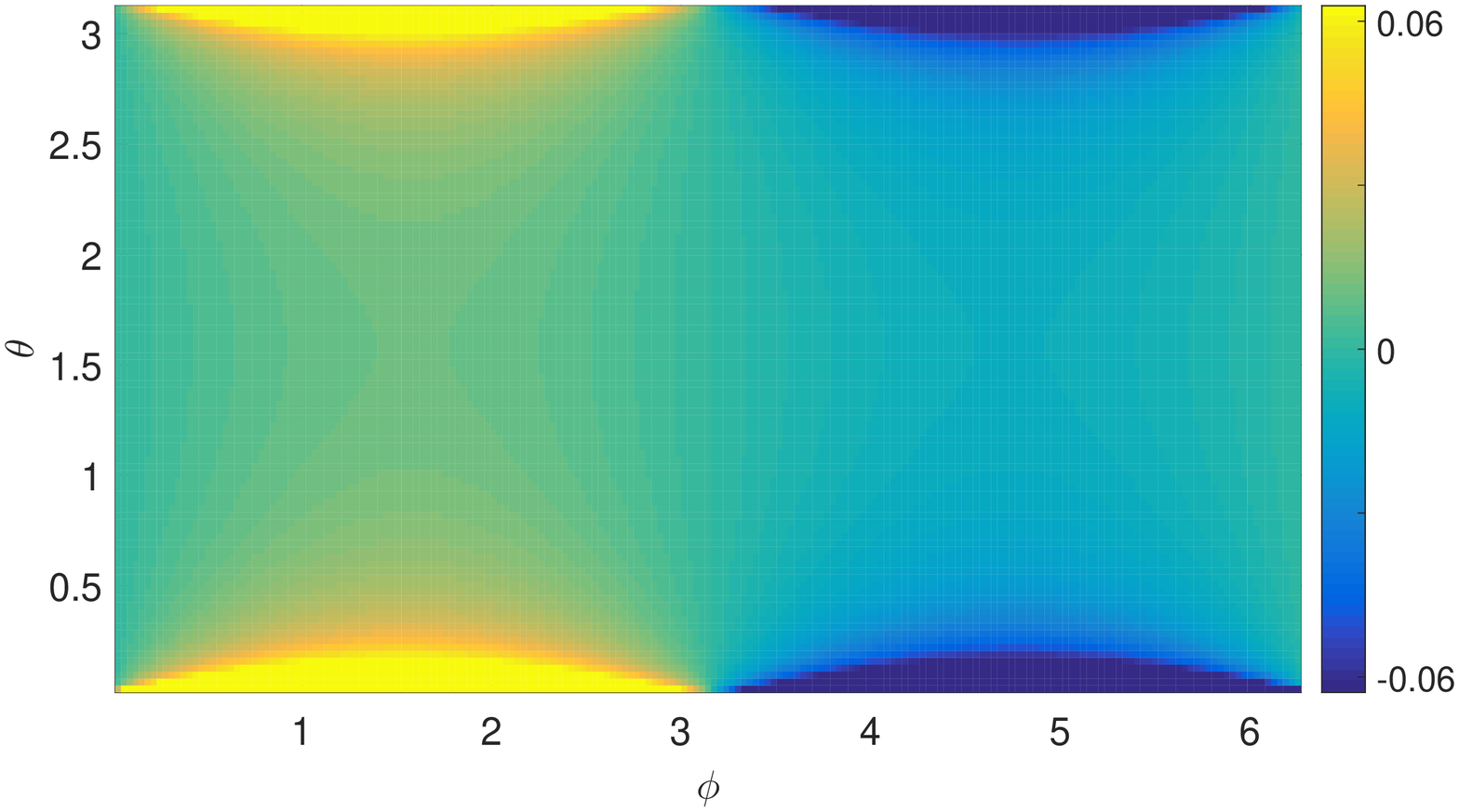}
\end{tabular}
\caption{The plot convention is the same to Fig.~\ref{fig1} but here the boost velocity is $\vec{v}=(0.01,0,0)$. Compared to Figs.~\ref{fig1}-\ref{fig3}, the velocity decreases and the angle changement decreases consequently. The color legend becomes smaller accordingly.}\label{fig4}
\end{figure}
In the last section we have assumed the boost velocity points to $z$ direction which results in $\lambda=0$. As an application of the Lorentz transformation formula (\ref{eq20}), we can calculate the phase change $\lambda$ due to boost for arbitrary velocity $\vec{v}$.

In order to calculate $\lambda$, we need $h\equiv h_+-ih_\times$ and $h'\equiv h'_+-ih'_\times$ which are related to
\begin{align}
h_{ij}&=h_+e^+_{ij}+h_\times e^\times_{ij},\\
h'_{ij}&=h'_+e'^+_{ij}+h'_\times e'^\times_{ij}.
\end{align}
But ones have to note that relation (\ref{eq28}) does not hold any more for general velocity $\vec{v}$. Keeping the above two relations in mind and multiplying $e^{ij}_+$ to the two sides of (\ref{eq20}), we get
\begin{align}
&h'_+e'^+_{ij}e^{ij}_++h'_\times e'^\times_{ij}e^{ij}_+=2h_++\frac{v^kh_{kl}v^l}{(1-\hat{r}\cdot\vec{v})^2}\frac{\gamma^2}{(1+\gamma)^2}v_iv_je^{ij}_+\nonumber\\
&-2\frac{v^kh_{kj}}{1-\hat{r}\cdot\vec{v}}\frac{\gamma}{1+\gamma}v_ie^{ij}_+.\label{eq31}
\end{align}

At general angular position $(\theta,\phi)$ we have relations
\begin{align}
\hat{\theta}^i&=\cos\theta\cos\phi\hat{e}_x+\cos\theta\sin\phi\hat{e}_y-\sin\theta\hat{e}_z,\\
\hat{\phi}^i&=-\sin\phi\hat{e}_x+\cos\phi\hat{e}_y.
\end{align}
Similarly we have
\begin{align}
\hat{\theta}'^i&=\cos\theta'\cos\phi'\hat{e}_x+\cos\theta'\sin\phi'\hat{e}_y-\sin\theta'\hat{e}_z,\\
\hat{\phi}'^i&=-\sin\phi'\hat{e}_x+\cos\phi'\hat{e}_y.
\end{align}
Here $(\theta,\phi)$ and $(\theta',\phi')$ are related through the aberration formula (\ref{eqII14}) which is equivalent to
\begin{align}
\hat{r}'&=\frac{\hat{r}\cdot\hat{v}-v}{1-\hat{r}\cdot\vec{v}}\hat{v}+\frac{1}{\gamma(1-\hat{r}\cdot\vec{v})}[\hat{r}-(\hat{r}\cdot\hat{v})\hat{v}],\\
\hat{r}\cdot\vec{v}&=v_x\sin\theta\cos\phi+v_y\sin\theta\sin\phi+v_z\cos\theta,\\
\hat{r}\cdot\hat{v}&=\frac{\hat{r}\cdot\vec{v}}{v}.
\end{align}
Compared to the aberration formula (\ref{eqII14}), the above expressions are more useful to the calculation here. These relations result in
\begin{align}
\cos\theta'&=\frac{1}{\gamma(1-\hat{r}\cdot\vec{v})}\left[\cos\theta-(\hat{r}\cdot\hat{v})\frac{v_z}{v}\right]\nonumber\\
&+\frac{\hat{r}\cdot\hat{v}-v}{1-\hat{r}\cdot\vec{v}}\frac{v_z}{v},\label{eq3}\\
\sin\theta'\cos\phi'&=\frac{1}{\gamma(1-\hat{r}\cdot\vec{v})}\left[\sin\theta\cos\phi-(\hat{r}\cdot\hat{v})\frac{v_x}{v}\right]\nonumber\\
&+\frac{\hat{r}\cdot\hat{v}-v}{1-\hat{r}\cdot\vec{v}}\frac{v_x}{v},\label{eq4}\\
\sin\theta'\sin\phi'&=\frac{1}{\gamma(1-\hat{r}\cdot\vec{v})}\left[\sin\theta\sin\phi-(\hat{r}\cdot\hat{v})\frac{v_y}{v}\right]\nonumber\\
&+\frac{\hat{r}\cdot\hat{v}-v}{1-\hat{r}\cdot\vec{v}}\frac{v_y}{v}.\label{eq13}
\end{align}
The combination of (\ref{eq4}) and (\ref{eq13}) can determine $\phi'$ in the range $(0,2\pi)$. According to the above relations, we can get explicitly
\begin{align}
&e'^+_{ij}e^{ij}_+=(\hat{\theta}\cdot\hat{\theta}')^2+(\hat{\phi}\cdot\hat{\phi}')^2-(\hat{\phi}\cdot\hat{\theta}')^2-(\hat{\theta}\cdot\hat{\phi}')^2,\\
&e'^\times_{ij}e^{ij}_+=2(\hat{\theta}'\cdot\hat{\theta})(\hat{\phi}'\cdot\hat{\theta})-2(\hat{\theta}'\cdot\hat{\phi})(\hat{\phi}'\cdot\hat{\phi}),\\
&h_{ij}v^iv^j=h_+e^+_{ij}v^iv^j+h_\times e^\times_{ij}v^iv^j,\\
&v^kh_{kj}v_ie^{ij}_+=h_+[(\hat{\theta}\cdot\vec{v})^2+(\hat{\phi}\cdot\vec{v})^2],\\
&e^+_{ij}v^iv^j=(\hat{\theta}\cdot\vec{v})^2-(\hat{\phi}\cdot\vec{v})^2,\\
&e^\times_{ij}v^iv^j=2(\hat{\theta}\cdot\vec{v})(\hat{\phi}\cdot\vec{v}).
\end{align}
Similar to (\ref{eq31}) we can use $e^{ij}_\times$ to multiply the two sides of (\ref{eq20}) and get
\begin{align}
&h'_+e'^+_{ij}e^{ij}_\times+h'_\times e'^\times_{ij}e^{ij}_\times=2h_\times+\frac{v^kh_{kl}v^l}{(1-\hat{r}\cdot\vec{v})^2}\frac{\gamma^2}{(1+\gamma)^2}v_iv_je^{ij}_\times\nonumber\\
&-2\frac{v^kh_{kj}}{1-\hat{r}\cdot\vec{v}}\frac{\gamma}{1+\gamma}v_ie^{ij}_\times,\label{eq32}
\end{align}
with
\begin{align}
&e'^+_{ij}e^{ij}_\times=2(\hat{\theta}'\cdot\hat{\theta})(\hat{\theta}'\cdot\hat{\phi})-2(\hat{\phi}'\cdot\hat{\theta})(\hat{\phi}'\cdot\hat{\phi}),\\
&e'^\times_{ij}e^{ij}_\times=2(\hat{\theta}'\cdot\hat{\theta})(\hat{\phi}'\cdot\hat{\phi})+2(\hat{\phi}'\cdot\hat{\theta})(\hat{\theta}'\cdot\hat{\phi}),\\
&v^kh_{kj}v_ie^{ij}_\times=h_\times[(\hat{\theta}\cdot\vec{v})^2+(\hat{\phi}\cdot\vec{v})^2].
\end{align}

Solving (\ref{eq31}) and (\ref{eq32}) for $h'_+$ and $h'_\times$ we get
\begin{align}
&h'_+=\frac{{\rm RHS}_1e'^\times_{ij}e^{ij}_\times-{\rm RHS}_2e'^\times_{ij}e^{ij}_+}{e'^+_{ij}e^{ij}_+e'^\times_{kl}e^{kl}_\times-e'^\times_{ij}e^{ij}_+e'^+_{kl}e^{kl}_\times},\label{eq5}\\
&h'_\times=\frac{{\rm RHS}_2e'^+_{ij}e^{ij}_+-{\rm RHS}_1 e_{ij}^{'+}e_{\times}^{ij}}{e'^+_{ij}e^{ij}_+e'^\times_{kl}e^{kl}_\times-e'^\times_{ij}e^{ij}_+e'^+_{kl}e^{kl}_\times},\label{eq6}
\end{align}
where ${\rm RHS}_{1,2}$ are respectively the right hand side of (\ref{eq31}) and (\ref{eq32}). Then $e^{2i\lambda}=\frac{h}{h'}$ gives us $\lambda(\theta,\phi,v_x,v_y,v_z)$.

As a self consistent check, we consider a special case $\vec{v}=v\hat{e}_z$ which should result in $\lambda(\theta,\phi,0,0,v_z)=0$. In this special case we have
\begin{align}
&\hat{r}\cdot\vec{v}=v\cos\theta,\\
&\hat{r}\cdot\hat{v}=\cos\theta,\\
&\cos\theta'=\frac{\cos\theta-v}{1-v\cos\theta},\sin\theta'=\frac{\sin\theta}{\gamma(1-v\cos\theta)},\\
&\phi'=\phi,\hat{\phi}'=\hat{\phi},\\
&\hat{\theta}\cdot\hat{\theta}'=1-\frac{v^2\sin^2\theta}{1-v\cos\theta}\frac{\gamma}{1+\gamma},\\
&\hat{\theta}'\cdot\hat{\phi}=\hat{\phi}'\cdot\hat{\theta}=0,\hat{\phi}'\cdot\hat{\phi}=1,\\
&e'^\times_{ij}e^{ij}_+=e'^+_{ij}e^{ij}_\times=0,\\
&e'^+_{ij}e^{ij}_+=1+(\hat{\theta}\cdot\hat{\theta}')^2,e'^\times_{ij}e^{ij}_\times=2(\hat{\theta}'\cdot\hat{\theta}),\\
&e^+_{ij}v^iv^j=v^2\sin^2\theta,e^\times_{ij}v^iv^j=0,\\
&h_{ij}v^iv^j=v^kh_{kj}v_ie^{ij}_+=h_+v^2\sin^2\theta,\\
&v^kh_{kj}v_ie^{ij}_\times=h_\times v^2\sin^2\theta,\\
&{\rm RHS}_1=h_+[1+(1-\frac{1}{1-v\cos\theta}\frac{\gamma}{1+\gamma}v^2\sin^2\theta)^2],\\
&{\rm RHS}_2=2h_\times(1-\frac{1}{1-v\cos\theta}\frac{\gamma}{1+\gamma}v^2\sin^2\theta).
\end{align}
In the calculation of $\hat{\theta}\cdot\hat{\theta}'$ we have used relation $1-\frac{1}{\gamma}=\frac{\gamma v^2}{1+\gamma}$. Based on the above calculation results we can get $h'=h$ which confirms $\lambda(\theta,\phi,0,0,v_z)=0$.

Formally Eqs.~(\ref{eq5}) and (\ref{eq6}) can be expressed as
\begin{align}
&h'_+=A_+h_++B_+h_\times,\\
&h'_\times=A_\times h_++B_\times h_\times,
\end{align}
where $A_{+,\times}$ and $B_{+,\times}$ only depend on $(\theta,\phi,v_x,v_y,v_z)$, or to say $A_{+,\times}$ and $B_{+,\times}$ are independent of $h_+$ and $h_\times$. Ones can verify that
\begin{align}
A_+=B_\times,A_\times=-B_+.
\end{align}
Consequently we have
\begin{align}
e^{-2i\lambda}=A_+-i A_\times,
\end{align}
which is independent of $h_+$ and $h_\times$. This is why we only write $\lambda(\theta,\phi,v_x,v_y,v_z)$ instead of $\lambda(\theta,\phi,v_x,v_y,v_z,h_+,h_\times)$. This property is also consistent to Eq.~(\ref{eq17}) which indicates that $\lambda$, as a Lorentz transformation factor, only depends on $(\theta,\phi,v_x,v_y,v_z)$.

Since $\lambda$ is independent of $h_+$ and $h_\times$, we can plug $h_+=1,h_\times=0$ into Eqs.~(\ref{eq5}) and (\ref{eq6}) to simplify the calculation of $\lambda$. Then we have
\begin{align}
&e^{-2i\lambda}=\nonumber\\
&\frac{\left({\rm F}_1e'^\times_{ij}e^{ij}_\times-{\rm F}_2e'^\times_{ij}e^{ij}_+\right)-i\left({\rm F}_2e'^+_{ij}e^{ij}_+-{\rm F}_1 e_{ij}^{'+}e_{\times}^{ij}\right)}{e'^+_{ij}e^{ij}_+e'^\times_{kl}e^{kl}_\times-e'^\times_{ij}e^{ij}_+e'^+_{kl}e^{kl}_\times},\label{eq15}\\
&{\rm F}_1\equiv2+(f_\theta-f_\phi)^2-2(f_\theta+f_\phi),\\
&{\rm F}_2\equiv2f_\theta f_\phi\frac{(\hat{\theta}\cdot\vec{v})^2-(\hat{\phi}\cdot\vec{v})^2}{(\hat{\theta}\cdot\vec{v})(\hat{\phi}\cdot\vec{v})},\\
&f_\theta\equiv\frac{\gamma}{1+\gamma}\frac{(\hat{\theta}\cdot\vec{v})^2}{1-\hat{r}\cdot\vec{v}},f_\phi\equiv\frac{\gamma}{1+\gamma}\frac{(\hat{\phi}\cdot\vec{v})^2}{1-\hat{r}\cdot\vec{v}}.
\end{align}

A time independent (equivalently frequency independent) phase factor $\lambda$ can be absorbed in the initial phase during the gravitational wave data analysis \cite{maggiore2008gravitational,creighton2012gravitational}.

As examples we plot the phase change and the wave propagating direction change in Figs.~\ref{fig1}-\ref{fig3}. Fig.~\ref{fig1} and Fig.~\ref{fig2} correspond to representative velocities $\vec{v}=0.9\hat{e}_x$ and $\vec{v}=0.9\hat{e}_y$ respectively. Fig.~\ref{fig3} corresponds to an arbitrary velocity $\vec{v}=(0.1143,0.8219,0.3481)$. There are some unsmooth places in the plots due to the range $(0,2\pi)$ taken by angles. The values of an angle $0$ and $2\pi$ are essentially the same.

Reminding that kick velocity of binary black hole (BBH) merger is about one percentage of the speed of light \cite{PhysRevLett.98.091101,PhysRevLett.98.231101,PhysRevLett.98.231102,2007ApJ...659L...5C,PhysRevLett.107.231102,PhysRevLett.117.011101,PhysRevLett.121.191102,PhysRevD.100.104039,2022arXiv220101302V}, we plot the results for $\vec{v}=0.01\hat{e}_x$ which can be compared to the high speed case shown in Fig.~\ref{fig1}. In Figs.~\ref{fig1}-\ref{fig3}, the involved velocity is order one and the corresponding angle changement is also order one.
In comparison, the involved velocity of Fig.~\ref{fig4} decreases to order $10^{-2}$ and the corresponding angle changement also decreases to order $10^{-2}$.
\section{Gravitational waveform of a moving BBH}\label{sec1}
In usual literature, ones use `detector frame' to mean coordinate system which moves along the detector and whose original point locates at the detector. Correspondingly `source frame' means the coordinate system which moves along the GW source and whose original point locates at the source. In the current section, there will be four different coordinate systems involved. Consequently we use `detector frame' and `source frame' to only mean the coordinate moving along the detector and source respectively. For both `detector frame' and `source frame', the original point of the coordinate system may locate at the detector or the source.
\begin{figure*}
\begin{tabular}{c}
\includegraphics[width=\textwidth]{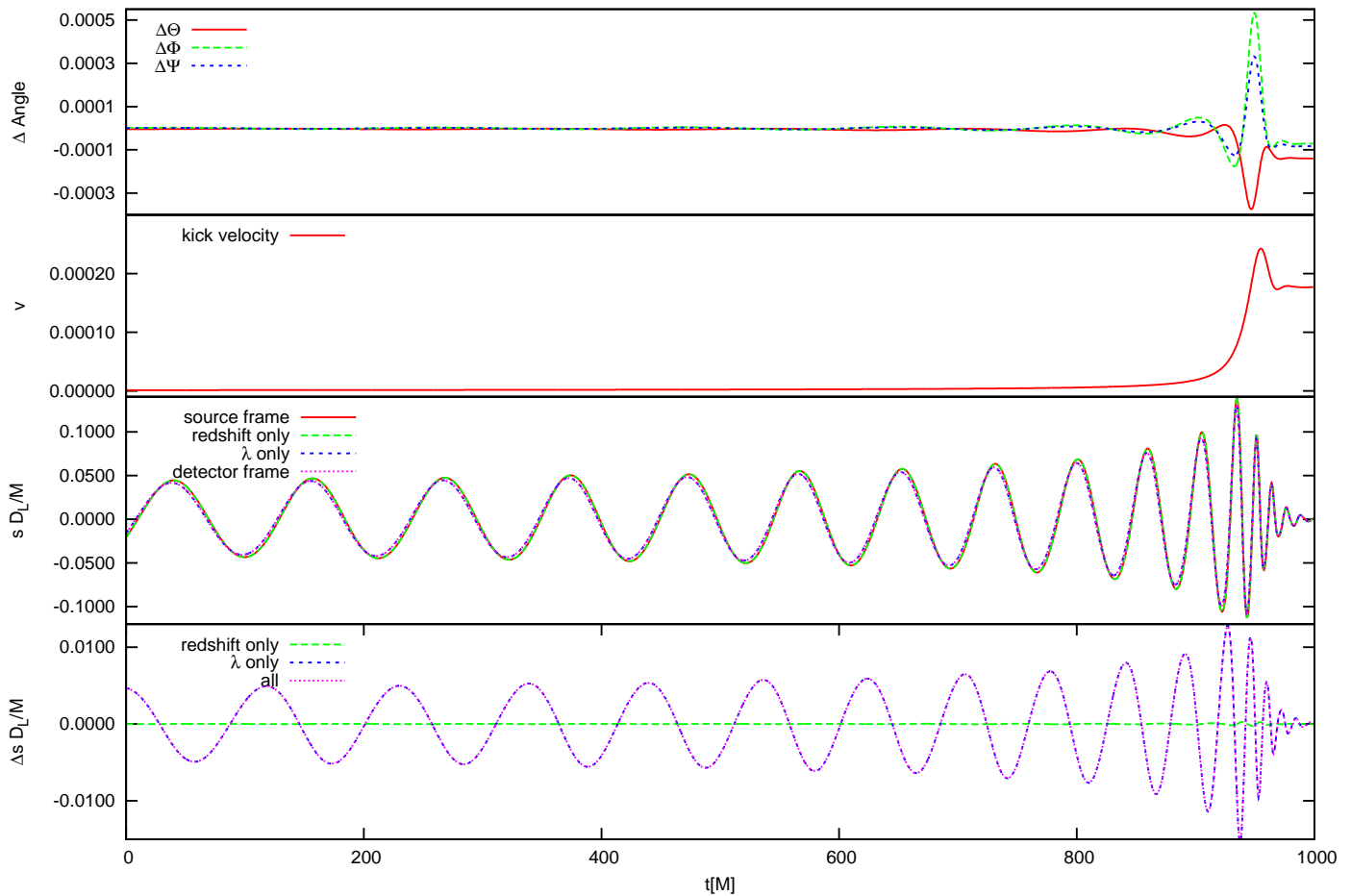}
\end{tabular}
\caption{The top panel: the source localization change $\Delta \Theta\equiv\Theta'-\Theta$, $\Delta \Phi\equiv\Phi'-\Phi$ and polarization angle change $\Delta \Psi\equiv\Psi'-\Psi$ of GW150914-like source due to the kick velocity through (\ref{eq9}), (\ref{eq10}) and (\ref{eq11}). The parameters of GW150914 are respectively symmetric mass ratio $\eta=0.248735$, effective spin $\chi_{s}=-0.303448,\chi_{a}=0.014667$, source localization $(\Theta,\Phi)=(2.77766633,1.6391)$, the polarization angle $\Psi=1.56749$ and inclination angle $\iota=2.6$ \cite{2021arXiv211103606T}. The second panel is for the corresponding kick velocity $v\equiv\sqrt{v_x^2+v_y^2+v_z^2}$. The third panel shows the the waveform of GW150914-like source due to the kick velocity. ``Source frame" means the waveform of rest source. ``Detector frame" means the waveform adjusted by the kick velocity. ``Redshift only" means adjustment only comes from the red shift factor $k$. ``$\lambda$ only" means the adjustment due to the Lorentz transformation but ignore the time difference between $t$ and $t'$. The bottom panel shows the waveform difference between adjusted ones and the source frame waveform. ``All" means the comparison between source frame waveform and detector frame waveform.}\label{fig5}
\end{figure*}

\begin{figure*}
\begin{tabular}{c}
\includegraphics[width=\textwidth]{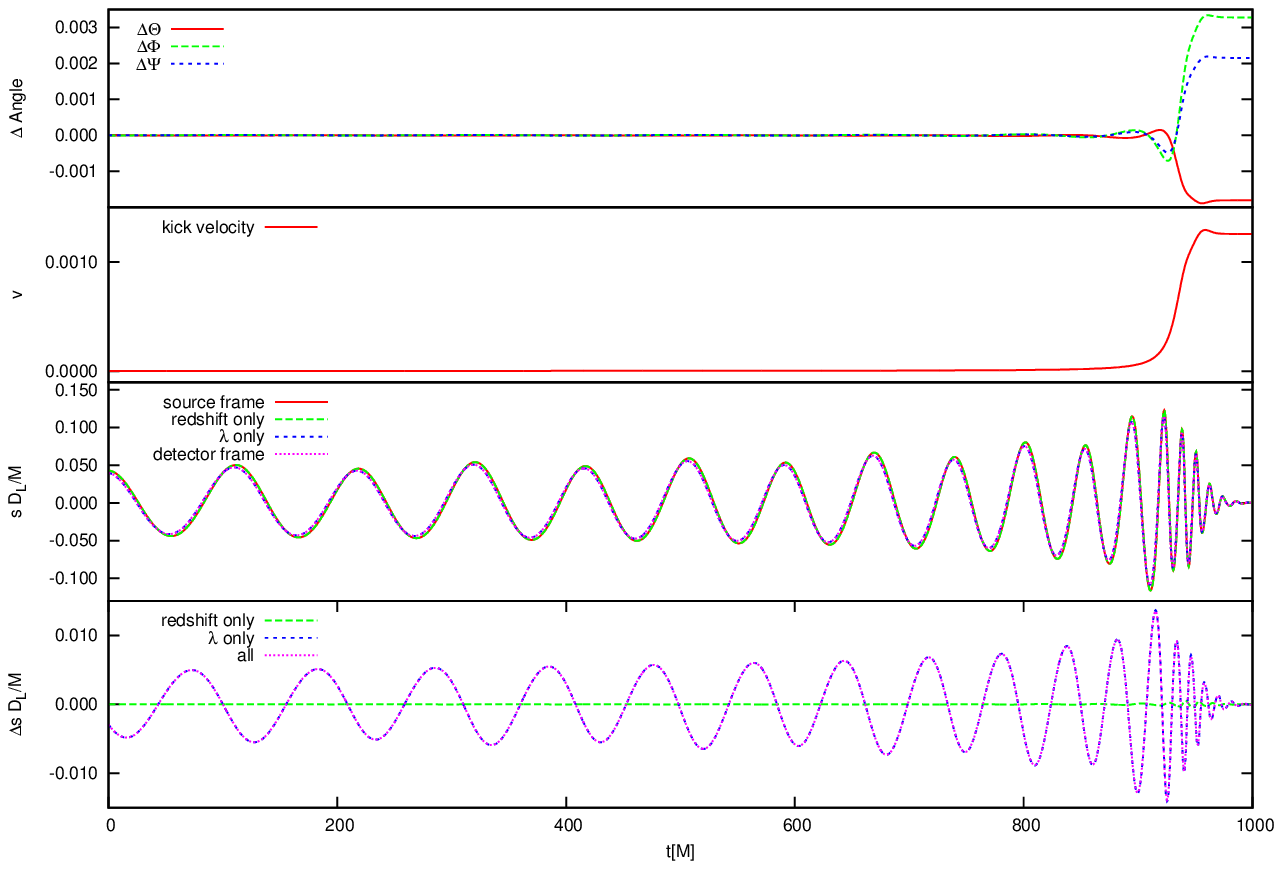}
\end{tabular}
\caption{The plot convention is the same to Fig.~\ref{fig5} but the intrinsic parameters of BBH are symmetric mass ratio $\eta=0.25$, effective spin $\chi_{s}=0,\chi_{a}=0.99$.}\label{fig6}
\end{figure*}

In the viewpoint of the source, no matter the detector is moving or not, the gravitational wave radiated in the same direction will be detected. As usual we choose a coordinate whose $z$ direction pointing to the orbital angular momentum of the BBH, and the $x$-$z$ plane contains the line connecting the BBH and the detector. Then the gravitational wave tensor reaching the detector $h_{ij}$ is determined by the intrinsic parameters of the BBH, the luminosity distance between the BBH and the detector $D_L$ and the inclination angle $\iota$. More specifically if the spin weighted -2 spherical harmonic components $h_{lm}(t;m_1,m_2,\vec{s}_1,\vec{s}_2)$ are given \cite{PhysRevD.96.044028_SEOBNRE,PhysRevD.101.044049_validSEOBNRE,PhysRevD.101.101501_TEOBeccc,Liu_2022,PhysRevD.105.044035}, we have
\begin{align}
h_+&=\frac{1}{D_L}\Re[\sum_{lm}h_{lm}(t;m_1,m_2,\vec{s}_1,\vec{s}_2){}_{-2}Y_{lm}(\iota,0)],\\
h_\times&=\frac{1}{D_L}\Im[\sum_{lm}h_{lm}(t;m_1,m_2,\vec{s}_1,\vec{s}_2){}_{-2}Y_{lm}(\iota,0)],\\
h_{ij}&=h_+e^+_{ij}(\iota,0)+h_\times e^\times_{ij}(\iota,0).
\end{align}
Here $m_{1,2}$ and $\vec{s}_{1,2}$ denote the mass and the spin of the two black holes as usual, ${}_{-2}Y_{lm}$ are the spin weighted -2 spherical harmonic functions.

Then changing to the viewpoint of the detector, Eq.~(\ref{eq20}) results in the needed gravitational wave tensor
\begin{align}
&h'_{ij}(t;m_1,m_2,\vec{s}_1,\vec{s}_2,\iota)=h'_{ij}(\frac{t'}{k};m_1,m_2,\vec{s}_1,\vec{s}_2,\iota),\\
&k=\frac{1}{\gamma(1-\vec{v}\cdot\hat{r})},\label{eq1}
\end{align}
where $t$ and $t'$ correspond to the time in the viewpoint of source frame and detector frame respectively.

Putting the coordinate origin at the detector, we get a spherical coordinate $(R,\Theta,\Phi)$ within the rest frame relative to the GW source (source frame). Within the moving frame respect to the GW source we denote the spherical coordinate as $(R',\Theta',\Phi')$ (detector frame). Assuming a GW source locates in direction $(\Theta,\Phi)$, the GW tensor can also be expanded as
\begin{align}
&h_{ij}=H_+E^+_{ij}+H_\times E^\times_{ij},\\
&E^+_{ij}=\hat{\Theta}_i\hat{\Theta}_j-\hat{\Phi}_i\hat{\Phi}_j,\\
&E^\times_{ij}=\hat{\Theta}_i\hat{\Phi}_j+\hat{\Theta}_j\hat{\Phi}_i
\end{align}

In the source frame, the bases $e^{+,\times}_{ij}$ and bases $E^{+,\times}_{ij}$ locate in the same plane. However, they may be different up to a rotation angle which is called the polarization angle $\Psi$
\begin{align}
e^+_{ij}&=\cos2\Psi E^+_{ij}-\sin2\Psi E^\times_{ij},\\
e^\times_{ij}&=\sin2\Psi E^+_{ij}+\cos2\Psi E^\times_{ij}.
\end{align}
$\Psi$ corresponds to the angle between $\hat{\theta}$ and $\hat{\Theta}$. Consequently we have
\begin{align}
h_{ij}&=h_+e^+_{ij}+h_\times e^\times_{ij}\nonumber\\
&=(h_+\cos2\Psi+h_\times\sin2\Psi)E^+_{ij}\nonumber\\
&+(h_\times\cos2\Psi-h_+\sin2\Psi)E^\times_{ij},\label{eq27}\\
H_+&=h_+\cos2\Psi+h_\times\sin2\Psi,\\
H_\times&=h_\times\cos2\Psi-h_+\sin2\Psi.
\end{align}
Similarly in detector frame we have
\begin{align}
H'_+&=h'_+\cos2\Psi'+h'_\times\sin2\Psi',\\
H'_\times&=h'_\times\cos2\Psi'-h'_+\sin2\Psi'.
\end{align}

According to the convention of traditional data analysis in gravitational wave detection, the waveform $h'_+$ and $h'_\times$ together with parameters $(m_1,m_2,\vec{s}_1,\vec{s}_2,D_L,\iota,\Theta',\Phi',\Psi',t'_c,\phi'_c)$ will be considered \cite{PhysRevD.92.044034}. Here $t'_c$ and $\phi'_c$ correspond to the coalescence time and the GW phase at that time. No matter the BBH is moving or not parameters $(m_1,m_2,\vec{s}_1,\vec{s}_2,D_L,\iota)$ do not change. In contrast $(\Theta',\Phi',\Psi',t'_c,\phi'_c)$ depend on the moving speed $\vec{v}$ of the BBH.

If the parameters $(\Theta,\Phi,\Psi,t_c,\phi_c)$ for the corresponding rest BBH are known, we can express $(\Theta',\Phi',\Psi',t'_c,\phi'_c)$ as functions of $(\Theta,\Phi,\Psi,t_c,\phi_c,\iota)$ and $\vec{v}$. Among these functions, $t'_c$ and $\phi'_c$ can be easily obtained as
\begin{align}
&t'_c=kt_c,\label{eq7}\\
&\phi'_c=\phi_c-2\lambda(\iota,0,v_x,v_y,v_z).\label{eq8}
\end{align}

Noting the gravitational wave propagates along $\hat{r}=-\hat{R}$ and $\hat{r}'=-\hat{R}'$ in the viewpoint of source frame and detector frame respectively, the aberration formula (\ref{eqII14}) tells us
\begin{align}
\cos\Theta'&=\frac{1}{\gamma(1+\hat{R}\cdot\vec{v})}\left[\cos\Theta+(\hat{R}\cdot\hat{v})\frac{v_Z}{v}\right]\nonumber\\
&-\frac{v+\hat{R}\cdot\hat{v}}{1+\hat{R}\cdot\vec{v}}\frac{v_Z}{v},\label{eq9}\\
\sin\Theta'\cos\Phi'&=\frac{1}{\gamma(1+\hat{R}\cdot\vec{v})}\left[\sin\Theta\cos\Phi-(\hat{R}\cdot\hat{v})\frac{v_X}{v}\right]\nonumber\\
&+\frac{v+\hat{R}\cdot\hat{v}}{1+\hat{R}\cdot\vec{v}}\frac{v_X}{v}.\label{eq10}
\end{align}
Similar to the notation for spherical coordinate, the unprimed notation $(X,Y,Z)$ means the Cartesian coordinate in source frame and $(X',Y',Z')$ means the one in detector frame. But different to the spherical coordinate $(R,\Theta,\Phi)$, the original point of the Cartesian coordinate $(X,Y,Z)$ locates at the detector.

In order to find $\Psi'$, we use the following steps. Since the waveform template based on the source frame is known we have $h=h_+-ih_\times$. Then we have
\begin{align}
&H_+-iH_\times\equiv H=he^{2i\Psi}.
\end{align}
The function  $\lambda$ solved in the last section can be used to calculate $H'$ and $h'$
\begin{align}
&H'_+-iH'_\times\equiv H'=He^{-2i\lambda(\Theta,\Phi,v_X,v_Y,v_Z)},\\
&h'_+-ih'_\times\equiv h'=he^{-2i\lambda(\theta,\phi,v_x,v_y,v_z)},
\end{align}
which leads to
\begin{align}
&H'=h'e^{2i\Psi'},\\
&\Psi'=\Psi-\lambda(\Theta,\Phi,v_X,v_Y,v_Z)+\lambda(\theta,\phi,v_x,v_y,v_z).\label{eq11}
\end{align}

Note more that the waveform model convention for source frame has $\theta=\iota,\phi=0$, we can relate $v_x,v_y,v_z$ to $\Theta,\Phi,v_X,v_Y,v_Z$ and $\iota$. Bases $(\hat{e}_X,\hat{e}_Y,\hat{e}_Z)$ and $(\hat{e}_x,\hat{e}_y,\hat{e}_z)$ are related through a rotation with Euler angles $(\Phi,\iota-\Theta,0)$. Consequently we have
\begin{align}
&\begin{pmatrix}v_x\\v_y\\v_z\end{pmatrix}=\begin{pmatrix}1&0&0\\0&\cos(\iota-\Theta)&\sin(\iota-\Theta)\\0&-\sin(\iota-\Theta)&\cos(\iota-\Theta)\end{pmatrix}\cdot\nonumber\\
&\begin{pmatrix}\cos\Phi&\sin\Phi&0\\-\sin\Phi&\cos\Phi&0\\0&0&1\end{pmatrix}
\begin{pmatrix}v_X\\v_Y\\v_Z\end{pmatrix}\\
&=\begin{pmatrix}\cos\Phi&
\sin\Phi&0\\
-\sin\Phi\cos(\iota-\Theta)&
\cos\Phi\cos(\iota-\Theta)&\sin(\iota-\Theta)\\
\sin\Phi\sin(\iota-\Theta)&-\cos\Phi\sin(\iota-\Theta)&\cos(\iota-\Theta)
\end{pmatrix}\nonumber\\
&\cdot \begin{pmatrix}v_X\\v_Y\\v_Z\end{pmatrix}. \label{eq2}
\end{align}

In conclusion we have got the waveform model for moving BBH
\begin{align}
&h'=h(\frac{t'}{k};m_1,m_2,\vec{s}_1,\vec{s}_2,\iota)e^{-2i\lambda(\iota,0,v_x,v_y,v_z)},\label{eq12}
\end{align}
where $v_x$, $v_y$, $v_z$ are functions of $\Theta$, $\Phi$, $\iota$, $v_X$, $v_Y$ and $v_Z$ as shown in (\ref{eq2}).

When the GW source is steadily moving, the velocity and consequently the phase factor are time independent. In this case, the phase factor $\lambda(\iota,0,v_x,v_y,v_z)$ can be absorbed into the parameter initial phase. And we can simplify the waveform template as
\begin{align}
h(\frac{t'}{k};m_1,m_2,\vec{s}_1,\vec{s}_2,\iota)
\end{align}
together with parameters $m_1$, $m_2$, $\vec{s}_1$, $\vec{s}_2$, $D_L$, $\iota$, $\Theta$, $\Phi$, $\Psi$, $t_c$, $\phi_c$, $v_X$, $v_Y$ and $v_Z$. However, we need to note that the parameters $(\iota,\Theta,\Phi,\Psi,t_c,\phi_c,v_X,v_Y,v_Z)$ are degenerated into parameters $(\iota,\Theta',\Phi',\Psi',t'_c,\phi'_c)$ according to the relations (\ref{eq7})-(\ref{eq10}), (\ref{eq11}) and (\ref{eq2}).

In contrast, if the GW source is accelerating \cite{PhysRevLett.98.091101,PhysRevLett.98.231101,PhysRevLett.98.231102,2007ApJ...659L...5C,PhysRevLett.107.231102,PhysRevLett.117.011101,PhysRevLett.121.191102,PhysRevD.100.104039,2022arXiv220101302V} the phase factor $\lambda(\iota,0,v_x,v_y,v_z)$ will depend on time and can not be absorbed into the parameter initial phase any more. Then our waveform model (\ref{eq12}) should be adopted. Consequently the initial phase is simplified to $\phi'_c=\phi_c$. However, parameters $\Theta'$, $\Phi'$ and $\Psi'$ will change with time.

In this accelerating case, the red shift factor $k$ will also depend on time. Consequently the relation between detector frame time $t'$ and the source frame time $t$ becomes
\begin{align}
t'=\int_0^t k(t) dt.
\end{align}

In Fig.~\ref{fig5} we use GW150914 \cite{PhysRevLett.116.061102} as an example to show the corresponding waveform $h'_+$, $h'_\times$ and the time dependence of parameters $\Theta'$, $\Phi'$ and $\Psi'$. In this example, the velocity $(v_x,v_y,v_z)$ corresponds to the kick velocity of the BBH due to the asymmetric gravitational radiation. Such kick velocity can be modeled through waveform model for a rest BBH. Then velocity $(v_X,v_Y,v_Z)$ can be obtained through the inverse transformation of (\ref{eq2})
\begin{align}
&\begin{pmatrix}v_X\\v_Y\\v_Z\end{pmatrix}=\nonumber\\
&\begin{pmatrix}\cos\Phi&
-\sin\Phi\cos(\iota-\Theta)&\sin\Phi\sin(\iota-\Theta)\\
\sin\Phi&
\cos\Phi\cos(\iota-\Theta)&-\cos\Phi\sin(\iota-\Theta)\\
0&\sin(\iota-\Theta)&\cos(\iota-\Theta)
\end{pmatrix}\nonumber\\
&\cdot\begin{pmatrix}v_x\\v_y\\v_z\end{pmatrix}.
\end{align}

The parameters of GW150914 are respectively total mass $M=65.677$M${}_\odot$, symmetric mass ratio $\eta=0.248735$, effective spin $\chi_{s}=-0.303448,\chi_{a}=0.014667$, luminosity distance $D_L=420$Mpc, source localization $(\Theta,\Phi)=(2.77766633,1.6391)$, the polarization angle $\Psi=1.56749$ and inclination angle $\iota=2.6$. These values are based on the posterior distribution data of these parameters, which is publicly available on the webpage of the LIGO open science center \cite{gw-openscience}. Here we have adopted the median values of the posterior distribution for the corresponding parameters. In the first panel of Fig.~\ref{fig5} we plot the source localization change $\Delta \Theta\equiv\Theta'-\Theta$, $\Delta \Phi\equiv\Phi'-\Phi$ and polarization angle change $\Delta \Psi\equiv\Psi'-\Psi$. The corresponding kick velocity is plotted in the second panel of Fig.~\ref{fig5}. The kick velocity is calculated based on gravitational waveform and the waveform is got through \texttt{SEOBNREHM} \cite{Liu_2022}. The source localization change and polarization angle change are much smaller than the measurement accuracy of current gravitational wave detection. In this case the kick velocity admits $10^{-4}$ of speed of light.

The maximal kick velocity of BBH is about $10^{-3}$ of speed of light \cite{PhysRevLett.98.091101,PhysRevLett.98.231101,PhysRevLett.98.231102,2007ApJ...659L...5C,PhysRevLett.107.231102,PhysRevLett.117.011101,PhysRevLett.121.191102,PhysRevD.100.104039,2022arXiv220101302V}. As an example we investigate the equal mass, anti-aligned spinning BBH with spin $\chi=0.99$. In Fig.~\ref{fig6} we plot the result. We find that the source localization change and polarization angle change increase one order compared to that of Fig.~\ref{fig5}. As the first detected GW event, GW150914 is a representative astrophysical source. Regarding to the waveform transformation we concern here, kick velocity is the only involved issue. Highest kick velocity of non-precessing BBHs corresponds to equal mass BBH with large spin component $\chi_a$ which is analyzed in Fig.~\ref{fig6}. Compared to such high kick velocity about $10^{-3}$ speed of light, mass ratio factor can only result in $10^{-4}$ speed of light at most \cite{PhysRevLett.98.091101}.

\begin{figure}
\begin{tabular}{c}
\includegraphics[width=0.45\textwidth]{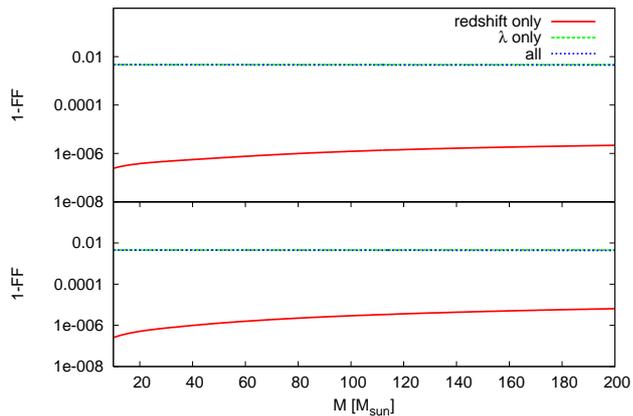}
\end{tabular}
\caption{The matching factor (FF) between the adjusted waveform due to the kick velocity and the unadjusted one. The top panel corresponds to the BBH shown in Fig.~\ref{fig5} and the bottom panel corresponds to the BBH shown in Fig.~\ref{fig6}.}\label{fig7}
\end{figure}

The GW strain detected by a detector can be described as \cite{maggiore2008gravitational,Liu_2022}
\begin{align}
s&=F^+h_++F^\times h_\times,\label{eq16}\\
F^+(\Theta,\Phi,\Psi)&\equiv\frac{1}{2}(1+\cos^2\Theta)\cos2\Phi\cos2\Psi\nonumber\\
&\,\,\,\,-\cos\Theta\sin2\Phi\sin2\Psi,\\
F^\times(\Theta,\Phi,\Psi)&\equiv\frac{1}{2}(1+\cos^2\Theta)\cos2\Phi\sin2\Psi\nonumber\\
&\,\,\,\,+\cos\Theta\sin2\Phi\cos2\Psi.
\end{align}
Note that the time dependence of $\Theta',\Phi',\Psi'$ will make the pattern functions $F^+$ and $F^\times$ vary with time also. In the third and fourth panels of Figs.~\ref{fig5} and \ref{fig6} we investigate the waveform change due to the kick velocity according to the Lorentz transformation. Changing from time $t$ to $t'$ corresponds to the red shift effect which has been studied in \cite{PhysRevLett.117.011101}. The combination of adjustment due to the phase factor $\lambda$, the source localization change and the polarization angle change is denoted as ``$\lambda$ only" in Figs.~\ref{fig5} and \ref{fig6}. This part is new compared to the study in \cite{PhysRevLett.117.011101}. However, we find that this part of changement is much larger than the red shift part. The waveform relative changing is about 10 percentage.

\begin{figure*}
\begin{tabular}{cc}
\includegraphics[width=0.45\textwidth]{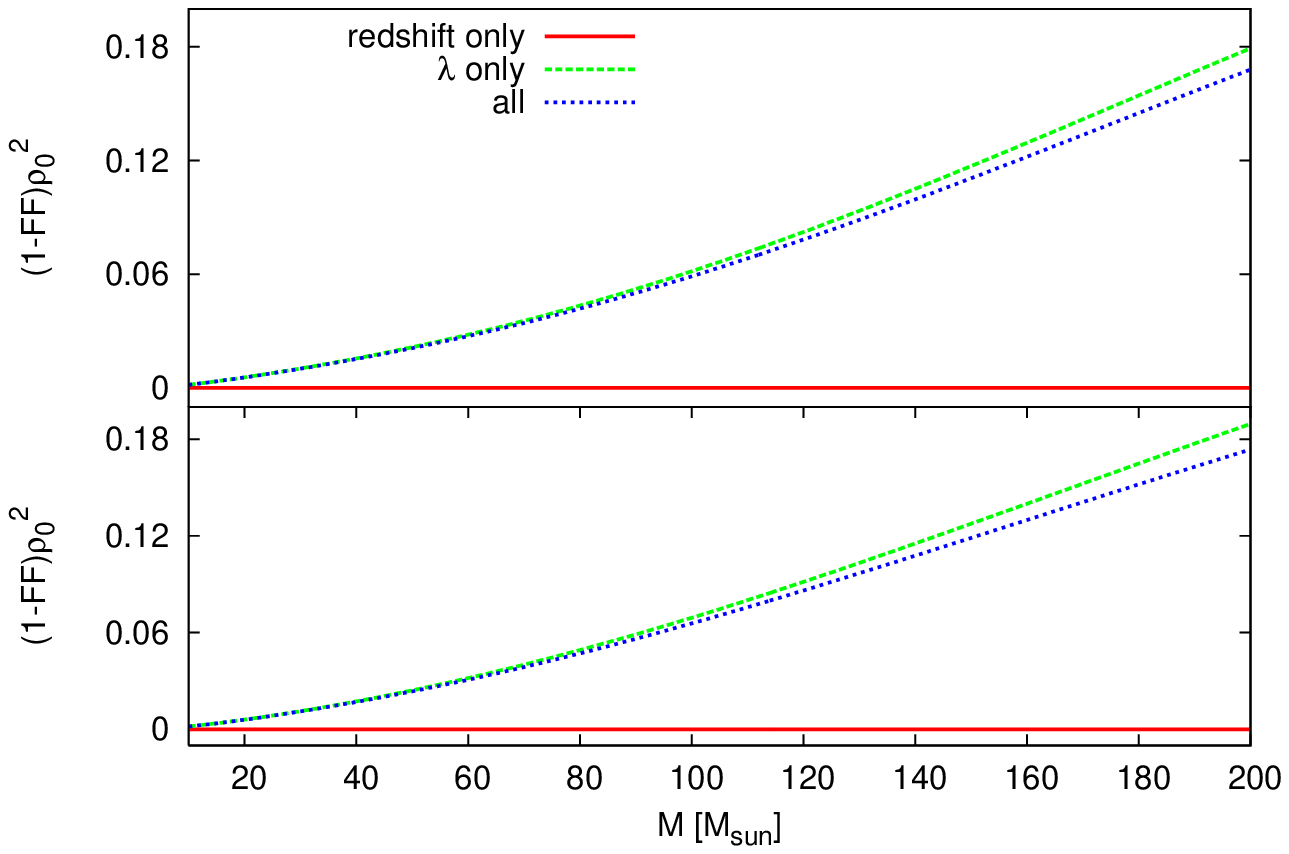}&
\includegraphics[width=0.45\textwidth]{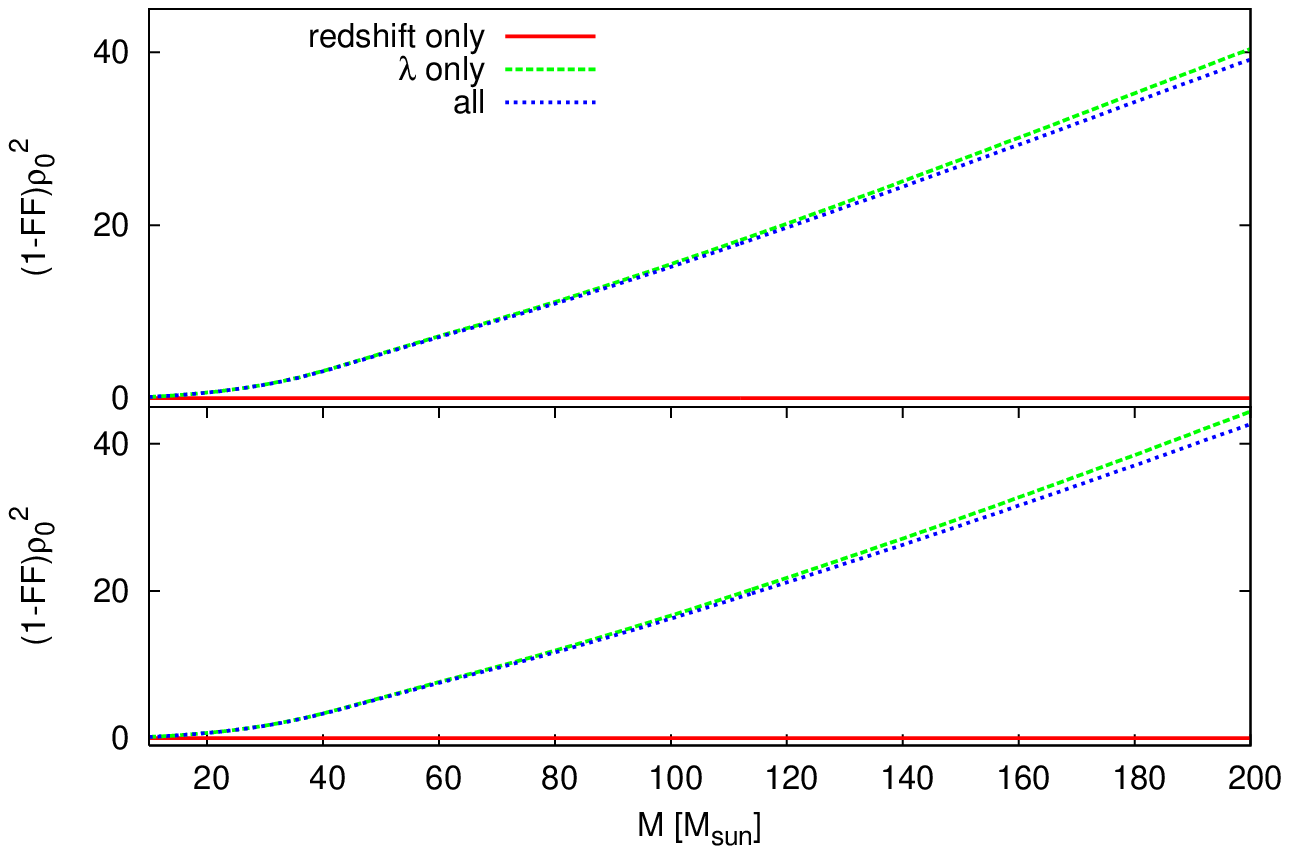}
\end{tabular}
\caption{The detectability criteria of the waveform correction introduced by the kick velocity of BBH involved in the Lorentz transformation waveform model. The left plot is for advanced LIGO sensitivity which corresponds to the following O4 observation run soon. The right plot is for the sensitivity of the Einstein Telescope (ET) which represents the 3G detector era in the near future. The top panels correspond to the BBH shown in Fig.~\ref{fig5} and the bottom panel corresponds to the BBH shown in Fig.~\ref{fig6}.}\label{fig8}
\end{figure*}
In order to quantify the waveform change, we calculate the matching factor respect to the designed sensitivity of advanced LIGO. We adopt the same procedure as we have done in \cite{Liu_2022}. The matching factor, also called faithfulness factor (FF), for two waveforms $s_1(t)$ and $s_2(t)$ is defined as
\begin{align}
\langle s_1|s_2\rangle &=4\mathcal{R}\int_{f_{\text{min}}}^{f_{\text{max}}} \frac{\tilde{s}_1(f) \tilde{s}_2^*(f)}{S_n(f)} df \nonumber \\
\text{FF} &\equiv \frac{\langle s_1|s_2\rangle }{\sqrt{\langle s_1|s_1\rangle \langle s_2|s_2\rangle  }},
\end{align}
where $S_n(f)$ is the one-sided power spectral density (PSD) of the detector noise, $(f_{\rm min},f_{\rm max})$ corresponds to the frequency range of the detector and the star notation ${-}^*$ means the complex conjugate. Corresponding to the mildly spinning GW150914-like BBH and the highly spinning anti-aligned BBH we plot the resulted matching factor in Fig.~\ref{fig7}. Both the designed sensitivity of advanced LIGO \cite{Sho10} and Einstein Telescope (ET) are calculated. The data ``LIGO-P1600143-v18-ET\_D.txt" \cite{lalET} are used for ET in the current work. Both results of advanced LIGO and ET are almost the same. Fig.~\ref{fig7} corresponds to the designed sensitivity of advanced LIGO. But the plot for ET is undistinguishable to that of Fig.~\ref{fig7}. If only the effect of red-shift is considered like \cite{PhysRevLett.117.011101}, the matching factors are larger than 99.999\%. When the $\lambda$ factor is taken into consideration, the matching factor decreases to about 99\%. Such high matching factor means the corrections introduced by the kick velocity can be neglected for current GW detectors \cite{PhysRevD.57.4566,PhysRevD.96.044028_SEOBNRE}.

In order to estimate the detectability of the correction introduced by the kick velocity involved in our waveform model, we investigate the following criteria \cite{PhysRevD.57.4566,PhysRevD.78.124020,PhysRevD.80.064019,PhysRevD.104.044037}
\begin{align}
&(1-{\rm FF})\rho_0^{2}\gtrsim1,\\
&\rho_0=\sqrt{\langle \delta h|\delta h\rangle},
\end{align}
where $\delta h$ means the waveform correction. The corresponding criteria are plotted in Fig.~\ref{fig8}. According to the above traditional criteria, we find that O4 may not be able to distinguish the difference of
the Lorentz transformation waveform model to the waveform without considering the effect of kick velocity. In contrast, the third generation (3G) detectors such as ET can well detect the effect of kick velocity. That is to say our waveform model
will be useful in 3G era.

When a BBH coalescence happens near a super-massive black hole \cite{PhysRevD.103.124044}, complicated accelerating process may appear. Such gravitational wave sources are called binary EMRI (Extreme Mass Ratio Inspirals) in Refs.~\cite{Chen2018,2019MNRAS.485L..29H,PhysRevLett.126.021101}. The above waveform construction process can be straightforwardly applied to set up the waveform template for the data analysis of binary EMRI systems.
\section{Summary and discussion}
The electric vector and the magnetic vector provide a good tool to describe electromagnetic field. That is because the electric vector and the magnetic vector present a traditional three dimensional picture which is easier to understand.

Similarly we have a three dimensional tensor for gravitational wave. Such a tensor provides a facility to transform between different coordinates. And also such a tensor makes people understand gravitational wave's behavior through traditional way instead of the complicated four dimensional object.

Unfortunately the gravitational wave tensor can only be used for three dimensional coordinate transformation. It can not be used to discuss the relation between two relatively moving observers. This is quite different to the electric vector and the magnetic vector. The electric vector and the magnetic vector are complete to describe electromagnetic field. Such completeness is due to the well known Lorentz transformation for the electric vector and the magnetic vector. Current paper filled this gap. We have constructed the wanted Lorentz transformation for gravitational wave tensor. Together with our Lorentz transformation rule (\ref{eq20}), we believe that the three dimensional tensor language will become more powerful to study gravitational wave physics and astronomy.

The Lorentz transformation for gravitational wave (\ref{eq20}) provides a good tool to construct theoretical waveform model for moving sources. Such waveform model will not be limited by small velocity approximation \cite{PhysRevD.100.063012,PhysRevD.101.083028,PhysRevLett.127.041102}. If only the GW waveform of a corresponding rest source is known, we can construct the three dimensional tensor and transform it to a moving frame for the source with any complicated motion \cite{PhysRevD.103.124044,PhysRevLett.127.041102}. Then the GW waveform can be straightforwardly reduced from the transformed three dimensional GW tensor.

Other than the waveform construction for moving sources, the Lorentz transformation for the three dimensional gravitational wave tensor can provide a powerful tool to study the interaction between a celestial body and a relatively moving GW source including the effect of GW on binary system \cite{PhysRevLett.128.101103,PhysRevD.105.064021}, the effect of GW on the relative motion between star and earth \cite{2018CQGra..35d5005K,PhysRevD.98.024020}, the effect of GW on the star seismic motion \cite{PAIK2009167,PhysRevD.90.102001,Harms_2021} and others.

In the viewpoint of BMS framework, gravitational wave appears in the order $(1/r)$ where $r$ is the area radius of the wave front. Alternatively in the viewpoint of flat spacetime perturbation, gravitational wave should be small. So we can conclude that $|h_{ij}|\ll1$ is required. It is interesting to ask whether this condition provides any limit for us when apply the Lorentz transformation rule (\ref{eq20}). In another word, is it possible that $|h_{ij}|\ll1$ while $|h'_{ij}|\geq1$ according to (\ref{eq20})?

Firstly the discussion after (\ref{eq20}) implies that
\begin{align}
&\frac{v^kh_{ki}}{1-\hat{r}\cdot\vec{v}}\sim v^kh_{ki},\\
&\frac{v^kh_{ki}v^i}{(1-\hat{r}\cdot\vec{v})^2}\sim v^kh_{ki}v^i.
\end{align}
And more we have
\begin{align}
&|\hat{r}_i|<1,\\
&\frac{\gamma}{1+\gamma}<1.
\end{align}
So the Lorentz transformation rule (\ref{eq20}) implies
\begin{align}
|h'_{ij}|&\sim|h_{ij}|+|v^kh_{ki}v^i|+2|v^kh_{ki}|\\
&\sim|h_{ij}|(1+2|v|+|v|^2)\\
&\sim|h_{ij}|.
\end{align}
This means the Lorentz transformation rule (\ref{eq20}) will preserve the smallness of the gravitational wave tensor. This fact makes sure that the Lorentz transformation rule (\ref{eq20}) is valid for all kinds of velocity $\vec{v}$.

Theoretical waveform model is important to gravitational wave data analysis. Current waveforms used by gravitational wave detection ignore the effect of moving velocity of the source relative to detector. In the current paper we calculated the explicit Lorentz transformation formula for the gravitational wave tensor, which is shown in (\ref{eq20}). This formula is a tensor equation. Any desired coordinate system can be used in specific application. Here we would like to emphasize that the Lorentz transformation formula (\ref{eq20}) is valid for arbitrary high velocity. There is no approximation involved in this formula.

The well known Bondi-Metzner-Sachs (BMS) transformation has already given out the Lorentz transformation of gravitational waveforms between two relatively moving frames. Such two waveforms are different up to a phase factor $\lambda$. But the phase factor $\lambda$ has not been explicitly calculated yet before. As an example of application of our formula (\ref{eq20}), we calculate straightforwardly the phase factor $\lambda$. Again our result (\ref{eq15}) is valid for arbitrary high velocity. There is no approximation involved in the calculation process.

If the gravitational wave source is moving with a constant velocity, the waveform transformation phase factor $\lambda$ from rest waveform is independent of time. Consequently such phase factor will completely degenerate with initial phase of the gravitational wave as shown in (\ref{eq12}). This means except the red shift factor $k$, no extra adjustment is needed for waveform template construction for moving sources. Correspondingly no information about the source velocity can be extracted by single detector. Only when two or more well separated detectors are available, our quantitative result (\ref{eq12}) and the aberration relation can be used to extract the information of the source velocity.

In contrast, if the gravitational wave source is accelerating, both the waveform transformation phase factor $\lambda$ and the aberration relations are time dependent which will contribute to the waveform. The combination of (\ref{eq9}), (\ref{eq10}) and (\ref{eq11}) together with (\ref{eq16}) is needed to construct waveform for moving sources. As an example we calculated the adjusted waveform by kick velocity of binary black hole merger. On the one hand our result indicates that such adjustment is ignorable for current gravitational wave detection but may be important to next generation detectors. On the other hand, this example shows that our construction procedure works well for waveform template construction of moving sources. Especially the binary EMRI sources may be a good application topic of our construction procedure. Regarding to the formation channel that BBH forms in a disk of a super-massive black hole, the waveform of moving sources will be important for BBHs locating nearer than $10^{6}$ gravitational radius to the center super-massive black hole in the near future 3G era.

\section*{Acknowledgments}
We thank Xian Chen, Alejandro Torres-Orjuela, Yun Fang, Kejia Lee and Lijing Shao for helpful discussions. This work was supported by CAS Project for Young Scientists in Basic Research YSBR-006, NSF of Hunan province (2018JJ2073) and the Key Project of Education Department of Hunan Province  (No. 21A0576).

\bibliographystyle{unsrt}
\bibliography{refs}

\end{document}